\documentclass[preprint,amsmath,amssymb]{revtex4}
\usepackage[latin1]{inputenc}
\usepackage{graphicx}
\usepackage[unicode=true,
 bookmarks=true,bookmarksnumbered=false,bookmarksopen=false,
 breaklinks=false,pdfborder={0 0 1},backref=false,colorlinks=false]
 {hyperref}

\makeatother

\begin{document}
\title{External sources in a minimal and nonminimal CPT-odd Lorentz violating Maxwell electrodynamics}

\author{L. H. C. Borges}
\email{luizhenriqueunifei@yahoo.com.br}

\author{A. F. Ferrari}
\email{alysson.ferrari@ufabc.edu.br}

\affiliation{Universidade Federal do ABC, Centro de Ci\^encias Naturais e Humanas,
Avenida dos Estados, 5001, 09210-580, Santo Andr\'e, SP, Brazil}

\begin{abstract}
This paper is devoted to the study of interactions between stationary
electromagnetic sources for the minimal and nonminimal CPT-odd photon
sector of the Standard Model Extension (SME), where we search mainly
for physical phenomena not present in the Maxwell electrodynamics.
First we consider the minimal CPT-odd sector, where the Lorentz violation
is caused by the Carroll-Field-Jackiw (CFJ) term, namely $\sim\epsilon^{\mu\nu\alpha\beta}\left(k_{AF}\right)_{\mu}A_{\nu}F_{\alpha\beta}$, and we treat the Lorentz breaking parameter $\left(k_{AF}\right)^{\mu}$
perturbatively up to second order. We consider effects due to the
presence of point-like charges, Dirac strings
and point-like dipoles. In special, we calculate the electromagnetic field
produced outside the string and investigate the so called Aharonov-Bohm
bound states in Lorentz violation context. After, we consider a model
where the Lorentz violation is generated by the higher-derivative
version of the CFJ model, namely $\sim\epsilon^{\mu\nu\alpha\beta}V_{\mu}A_{\nu}\Box F_{\alpha\beta}$,
which is a dimension five term of the CPT-odd sector of the nonminimal
SME. For this higher-derivative model, we obtain effects up to second
order in $V^{\mu}$ related to the presence of point-like charges
and a steady current line. We use overestimated constrains for the
Lorentz violation parameters in order to investigate the physical
relevance of some results found in atomic systems. 
We also make an overestimate for the background vectors using experimental data from the atomic electric field.
\end{abstract}

\maketitle

\section{Introduction\label{I}}

Lorentz violation theories have been intensively investigated in the
literature. The main reason for this arises from the fact that the
Lorentz symmetry could be violated at higher energy scales, where
quantum gravitational effects can not be neglected. Most of these
investigations have been performed in the framework of the Standard
Model Extension (SME) \cite{mSME1,mSME2,mSME3}, also called minimal SME. We mention for instance, investigations which concern the photon CPT-even sector \cite{LHCFABJHN,LHCFAB1,Plate1,Hydr,even1,even2,even3,
even4,even5,even6,even7,even8,even9,even10,even11,even12}, photon CPT-odd sector \cite{CFJ,odd2,odd3,odd4,odd5,odd6,odd7,odd8}, fermion sector \cite{fer1,fer2,fer3,fer4,fer5}, QED \cite{qed1,qed2,qed3,qed4,qed5}, and many others. 

In the latest years, Lorentz-violating terms containing higher derivatives have been studied in the literature. 
These terms are called nonminimal operators and belong to the nonminimal SME, which can be understood as an effective field theory that includes these non renormalizable operators.  These later can be relevant in the search for Lorentz violation signals in experiments involving particles of high energies and even of low energies, which is surprising since these nonminimal terms are expected to be suppressed in this energy scale.
The study of the nonminimal operators was started with the development of the nonminimal versions of the SME  for both the fermion \cite{ferNM1} and the photon sector \cite{phNM1}. We also mention investigations which concern the point-charge self-energy \cite{LHCFABAFF2}, the study of radiative corrections \cite{RC1,RC2,RC3,RC4,AltschulRC}, the search for Lorentz violation effects through boundary conditions \cite{BCNM}. Several other relevant aspects involving nonminimal operators can be found in the Refs. \cite{MMFCPTodd,VAKZL,MMFCPTeven,MM85,MS89,MS90,FABWave,
AF95,PRA791,YD94,LHCAFFFAB1}.

Regarding the interactions between external sources in a Lorentz violation scenario, we highlight studies in a minimal \cite{LHCFABJHN,LHCFAB1} and nonminimal \cite{LHCAFFFAB1} CPT-even photon sector of the SME. However, investigations of this type have not yet been carried out in the CPT-odd photon sector up to now. It could be of great interest to investigate what types of physical phenomena can arise in this sector due to such
interactions. In addition, issues related to this topic have been very little investigated in the context of the SME.

This paper is devoted to this subject where for both the minimal and nonminimal CPT-odd photon sector of the SME, we search mainly for effects produced by the presence of stationary field sources not present in the Maxwell
electrodynamics.

In the first part of the paper, we consider the minimal CPT-odd sector,
where the Maxwell Lagrangian is augmented by the Carroll-Field-Jackiw
(CFJ) term $\sim\epsilon^{\mu\nu\alpha\beta}\left(k_{AF}\right)_{\mu}A_{\nu}F_{\alpha\beta}$
\cite{CFJ}. Since the background vector $\left(k_{AF}\right)^{\mu}$
is assumedly very small we treat it perturbatively up to second order. Specifically,
we show the appearance of a spontaneous torque on a typical electromagnetic
dipole, as well as a magnetic field produced by a static point-like charge. We also consider some effects due
to the presence of Dirac strings. We show that the Dirac string interacts
with a point-like charge and with another Dirac
string, and we calculate the electromagnetic field configuration produced outside the string. We also obtain the corrections in the energy levels
of the 2-dimensional quantum rigid rotor, which concerns the Aharonov-Bohm
bound states. For completeness, we discuss the case of sources
which describe steady point-like dipoles.

In the second part of the paper, we consider a model where we added
in the Maxwell Lagrangian a dimension five term of the CPT-odd sector
of the nonminimal SME, namely, $\sim\epsilon^{\mu\nu\alpha\beta}V_{\mu}A_{\nu}\Box F_{\alpha\beta}$, which is the higher-derivative version of the CFJ
term. Some aspects involving this model were investigated in \cite{AltschulRC,MMFCPTodd,MMFwave}. We treat the Lorentz breaking parameter $V^{\mu}$ up to second
order, and show that the presence of point-like charges and a steady
current line induce nontrivial effects in low energy physics, as spontaneous
torques and electromagnetic fields.  

Along the paper, we use overestimated constrains for the
Lorentz violation parameters in order to investigate the physical relevance of some results found in atomic systems. 
We also make an overestimate for the background vectors using experimental data from the atomic electric field.

The paper is organized as follows: in section \ref{II} for the minimal
CPT-odd sector, we consider effects related to electromagnetic sources
describing point-like stationary charges,
Dirac strings and steady point-like dipoles. In section \ref{NM}
we obtain some physical phenomena due to the presence of point-like
charges and a steady current line for a specific model belonging
to the nonminimal CPT-odd sector of the SME. Section \ref{conclusoes}
is devoted to conclusions and final remarks.

In this paper we work in a $3+1$-dimensional Minkowski space-time
with metric $\eta^{\rho\nu}=(1,-1,-1,-1)$. The Levi-Civita tensor
is denoted by $\epsilon^{\rho\nu\alpha\beta}$ with $\epsilon^{0123}=1$.

\section{Minimal CPT-odd sector\label{II}}

The electromagnetic CPT-odd sector of the minimal SME is defined by
the following Lagrangian density \cite{mSME1}, 
\begin{equation}
{\cal L}=-\frac{1}{4}F_{\mu\nu}F^{\mu\nu}-\frac{1}{2\gamma}\left(\partial_{\mu}A^{\mu}\right)^{2}+\frac{1}{2}\epsilon^{\mu\nu\alpha\beta}\left(k_{AF}\right)_{\mu}A_{\nu}F_{\alpha\beta}+J^{\mu}A_{\mu}\ ,\label{lagEm}
\end{equation}
where $A^{\mu}$ is the photon field, $F^{\mu\nu}=\partial^{\mu}A^{\nu}-\partial^{\nu}A^{\mu}$
the corresponding field strength, $J^{\mu}$ is an external source
and $\gamma$ is the gauge parameter. Lorentz violation is introduced
by the constant background vector $\left(k_{AF}\right)^{\mu}$, which
has mass dimension one. Due to the complexity of the calculations,
the LV parameter will be treated perturbatively, since it is assumedly
a very small quantity compared to any relevant physical scale in the
problem.

By fixing the Feynman gauge $\gamma=1$ and using standard methods
of quantum field theory, one can show that the propagator for the
model (\ref{lagEm}) up to second order in the background vector reads
\begin{equation}
\label{propEm}
D^{\mu\nu}(x,y)=\int\frac{d^{4}p}{(2\pi)^{4}}\left[\Delta^{\mu\nu}\left(p\right)+\Delta_{LV}^{\mu\nu}\left(p\right)\right]e^{-ip\cdot\left(x-y\right)}\ ,  
\end{equation}
where we defined, in momentum space, the standard Maxwell propagator,
\begin{equation}
\Delta^{\mu\nu}\left(p\right)=-\frac{\eta^{\mu\nu}}{p^{2}}\ , \label{PropM}
\end{equation}
and the modification induced by Lorentz violation, reading 
\begin{align}
\label{propEM2}
\Delta_{LV}^{\mu\nu}\left(p\right) & =-\frac{1}{p^{2}}\Biggl\{\eta^{\mu\nu}\Biggl[\frac{4\left[\left(k_{AF}\right)\cdot p\right]^{2}}{p^{4}}-\frac{4\left(k_{AF}^{2}\right)}{p^{2}}\Biggr]+\frac{4\left(k_{AF}^{2}\right)}{p^{2}}\frac{p^{\mu}p^{\nu}}{p^{2}}+\frac{2i}{p^{2}}\epsilon^{\mu\nu\alpha\beta}\left(k_{AF}\right)_{\alpha}p_{\beta}\nonumber \\
 & +\frac{4\left(k_{AF}\right)^{\mu}\left(k_{AF}\right)^{\nu}}{p^{2}}-\frac{4\left[\left(k_{AF}\right)\cdot p\right]}{p^{4}}\left[\left(k_{AF}\right)^{\mu}p^{\nu}+\left(k_{AF}\right)^{\nu}p^{\mu}\right]\Biggr\}\ .
\end{align}

We notice that above propagator is in accordance with that one obtained in the appendix A of Ref. \cite{MMFCPTodd}. In order to evince this fact we must perform in Eq. (A4b) of this appendix the substitutions $\xi=-1$, $V_{\mu}\rightarrow -2\left(k_{AF}\right)_{\mu}$ and after we make an expansion up to second order in the background vector.

Since we have a quadratic Lagrangian in the gauge field, the contribution
to the vacuum energy of the system due to the stationary field source
$J^{\mu}\left(x\right)$ is given by \cite{LHCFABJHN,LHCAFFFAB1,LHCFAB1,FABGFH1,LHCFABAFF2}
\begin{eqnarray}
E & = & \frac{1}{2T}\int\int d^{4}x\ d^{4}yJ^{\mu}(x)D_{\mu\nu}(x,y)J^{\nu}(y)\nonumber \\
 & = & \frac{1}{2T}\int\frac{d^{4}p}{(2\pi)^{4}}\int\int d^{4}x\ d^{4}y\ e^{-ip\cdot\left(x-y\right)}J_{\mu}\left(x\right)\left[\Delta^{\mu\nu}\left(p\right)+\Delta_{LV}^{\mu\nu}\left(p\right)\right]J_{\nu}\left(y\right)\ ,\label{zxc1}
\end{eqnarray}
where $T$ stands formally for the fact that the spacetime integrals
are to be performed over a finite range of the temporal variable,
and the infinite time limit is to performed after the result is divided
by this range of time. From this equation, we can compute the interaction
energy between different electromagnetic field sources for the model
(\ref{lagEm}), as we will do for several interesting cases in the
following.

\subsection{Point-like charges\label{III}}

\subsubsection{Interaction energy and torque\label{ETorque}}

We start by considering the interaction between two steady point-like
charges. The corresponding external source reads 
\begin{equation}
J_{\mu}^{CC}({\bf x})=q_{1}\eta_{\mu}^{0}\delta^{3}\left({\bf x}-{\bf a}_{1}\right)+q_{2}\eta_{\mu}^{0}\delta^{3}\left({\bf x}-{\bf a}_{2}\right)\ ,\label{corre1Em}
\end{equation}
where the location of the charges $q_{1}$ and $q_{2}$ are given
by ${\bf a}_{1}$ and ${\bf a}_{2}$, and the label $CC$ is to remind
us we are looking for the charge-charge potential. Substituting (\ref{corre1Em})
in (\ref{zxc1}), discarding the self-interaction, and performing
the integrals in the following order: $d^{3}{\bf x}$, $d^{3}{\bf y}$,
$dx^{0}$, $dp^{0}$ and $dy^{0}$, we arrive at 
\begin{equation}
E^{CC}=q_{1}q_{2}\int\frac{d^{3}{\bf p}}{(2\pi)^{3}}\left[\Delta^{00}\left(p^{0}=0,{\bf {p}}\right)+\Delta_{LV}^{00}\left(p^{0}=0,{\bf {p}}\right)\right]e^{i{\bf p}\cdot{\bf a}}\ ,\label{Ener2EM}
\end{equation}
where we defined ${\bf {a}={\bf a}_{1}-{\bf a}_{2}}$, as the relative
position of the two charges. From Eqs. (\ref{PropM}) and \textbf{(\ref{propEM2})},
one can write 
\begin{equation}
E^{CC}=q_{1}q_{2}\left[\int\frac{d^{3}{\bf p}}{(2\pi)^{3}}\frac{e^{i{\bf p}\cdot{\bf a}}}{{\bf {p}}^{2}}-4\left({\bf {k}}_{AF}\right)^{2}\int\frac{d^{3}{\bf p}}{(2\pi)^{3}}\frac{e^{i{\bf p}\cdot{\bf a}}}{{\bf {p}}^{4}}-4\left[\left({\bf {k}}_{AF}\right)\cdot{\bf {\nabla}}_{{\bf {a}}}\right]^{2}\int\frac{d^{3}{\bf p}}{(2\pi)^{3}}\frac{e^{i{\bf p}\cdot{\bf a}}}{{\bf {p}}^{6}}\right]\ ,\label{ECC1}
\end{equation}
where ${\bf {\nabla}}_{{\bf {a}}}=\left(\frac{\partial}{\partial a^{1}},\frac{\partial}{\partial a^{2}},\frac{\partial}{\partial a^{3}}\right)$.
Using the fact that \cite{Gradshteyn} 
\begin{equation}
\int\frac{d^{3}{\bf p}}{(2\pi)^{3}}\frac{e^{i{\bf p}\cdot{\bf a}}}{{\bf {p}}^{2}}=\frac{1}{4\pi\mid{\bf {a}}\mid}\ ,\ \int\frac{d^{3}{\bf p}}{(2\pi)^{3}}\frac{e^{i{\bf p}\cdot{\bf a}}}{{\bf {p}}^{4}}=-\frac{\mid{\bf {a}}\mid}{8\pi}\ ,\ \int\frac{d^{3}{\bf p}}{(2\pi)^{3}}\frac{e^{i{\bf p}\cdot{\bf a}}}{{\bf {p}}^{6}}=\frac{\mid{\bf {a}}\mid^{3}}{96\pi}\ ,\label{INT}
\end{equation}
and carrying out some manipulations, we obtain 
\begin{equation}
E^{CC}=\frac{q_{1}q_{2}}{4\pi\mid{\bf {a}}\mid}\left[1+\frac{{\bf {a}}^{2}}{2}\left(3\left({\bf {k}}_{AF}\right)^{2}-\frac{\left[\left({\bf {k}}_{AF}\right)\cdot{\bf {a}}\right]^{2}}{{\bf {a}}^{2}}\right)\right]\ .\label{Ener6EM}
\end{equation}

The expression in (\ref{Ener6EM}) is a perturbative result up to
second order in the background LV vector for the interaction energy
between two stationary point-like charges for the model (\ref{lagEm}).
The first term between brackets in the right hand side of the Eq.
(\ref{Ener6EM}) corresponds to the well-known Coulombian interaction,
while the remaining terms are anisotropic corrections which account
for the Lorentz violation. In the setup where $\left(k_{AF}\right)^{\mu}=\left((k_{AF})^{0},0\right)$ the LV effects disappear.

\begin{figure}[!h]
\centering \includegraphics[scale=0.30]{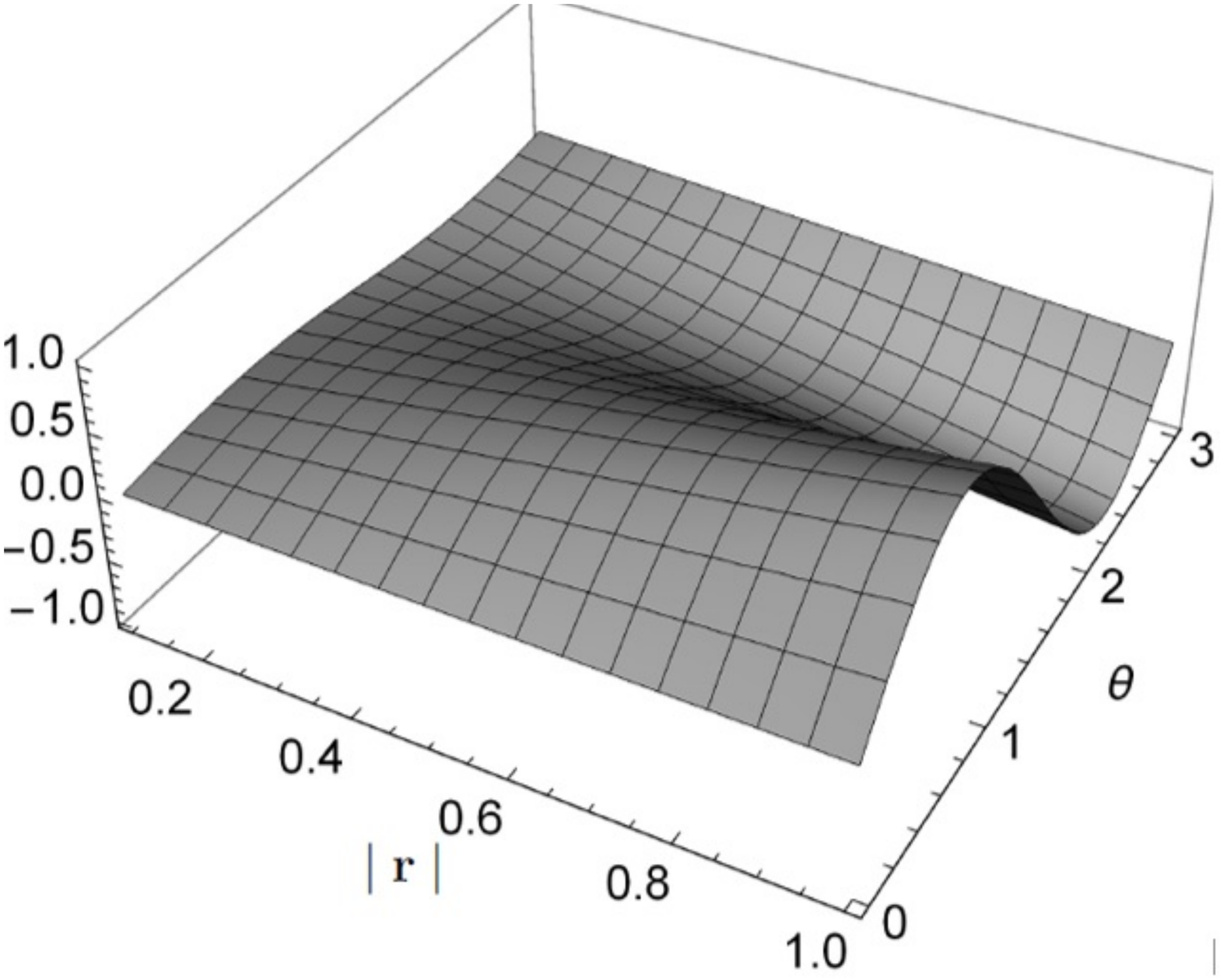} \caption{Torque (\ref{TorqueEM}) multiplied by $\frac{8\pi}{q^{2}\left({\bf{k}}_{AF}\right)^{2}}$.}
\label{grafico1}
\end{figure}

The nontrivial dependence of the energy (\ref{Ener6EM}) on the orientation of the background
vector with respect to ${\bf {a}}$ leads to a spontaneous torque
on an electric dipole, similarly to what was pointed out in Refs.\,\cite{LHCFABJHN,LHCAFFFAB1,LHCFAB1}.
In order to investigate this phenomena, we consider a typical dipole
composed of two opposite electric charges with $q_{1}=-q_{2}=q$ located
at positions ${\bf a}_{1}={\bf d}+\frac{{\bf r}}{2}$ and ${\bf a}_{2}={\bf d}-\frac{{\bf r}}{2}$,
with ${\bf r}$ taken to be a fixed vector. Therefore the Eq. (\ref{Ener6EM})
becomes 
\begin{equation}
E_{dipole}^{CC}=-\frac{q^{2}}{4\pi}\left[\frac{1}{\mid{\bf r}\mid}+\frac{\mid{\bf r}\mid\left({\bf {k}}_{AF}\right)^{2}}{2}\left[3-\cos^{2}(\theta)\right]\right]\ ,\label{Ener7EM}
\end{equation}
where $\theta$ is the angle between the vectors ${\bf r}$ and $\left({\bf {k}}_{AF}\right)$,
with $0\leq\theta\leq\pi$. The corresponding torque on the dipole
reads 
\begin{eqnarray}
\tau_{dipole}^{CC} & = & -\frac{\partial E_{dipole}^{CC}}{\partial\theta}=\frac{q^{2}\mid{\bf r}\mid\left({\bf {k}}_{AF}\right)^{2}}{8\pi}\sin(2\theta)\ .\label{TorqueEM}
\end{eqnarray}
This spontaneous torque on the dipole is an exclusive effect due to
the Lorentz symmetry breaking. It vanishes for the special orientations
$\theta=0,\pi/2,\pi$, as well for the obvious case in which the LV
is removed, and its general behavior can be seen in Fig.\,\ref{grafico1}.

The Lorentz violation parameter is strongly constrained from astrophysical
test data to be of order $\mid k_{AF}\mid\sim10^{-42}$GeV \cite{CFJ},
and we will use this upper bound to obtain order of magnitude estimates
for the torque (\ref{TorqueEM}). For this task, we consider a typical
microscopic system of condensed matter (atomic system), with distances
of order of angstroms, $|{\bf r}|\sim10^{-10}$m, electric charges
equal, in magnitude, to the electron's charge, $q\sim1.60217\times10^{-19}$C.
Substituting these in (\ref{TorqueEM}) we obtain $\tau_{dipole}^{CC}\sim10^{-91}$Nm. One of the possible ways to measure this type of torque is through torsion pendulum experiments where it is possible to reach values of order $10^{-16}$Nm \cite{torque}. Therefore, the obtained  effect are out of reach of being measured by using current technology.

\subsubsection{Electromagnetic field configuration\label{EMcharge}}

Here, we obtain the electromagnetic field configuration generated by a stationary point-like charge placed at origin in an arbitrary point ${\bf{x}}=\left(x^{1},x^{2},x^{3}\right)$. We start by considering the field configuration produced by a given
external source, which can be written as
\begin{equation}
A^{\mu}\left(x\right)=A_{M}^{\mu}\left(x\right)+\Delta A_{LV}^{\mu}\left(x\right)\ ,\label{fields2}
\end{equation}
with 
\begin{align}
A_{M}^{\mu}\left(x\right) & =\int\frac{d^{4}p}{(2\pi)^{4}}\int d^{4}y\ e^{-ip\cdot\left(x-y\right)}\Delta^{\mu\nu}\left(p\right)J_{\nu}(y)\ ,\label{defAs}\\
\Delta A_{LV}^{\mu}\left(x\right) & =\int\frac{d^{4}p}{(2\pi)^{4}}\int d^{4}y\ e^{-ip\cdot\left(x-y\right)}\Delta_{LV}^{\mu\nu}\left(p\right)J_{\nu}(y)\ ,\label{defalv}
\end{align}  
where the first contribution in Eq. (\ref{fields2}) faccounts for the standard Maxwell theory, and the second one is due to the Lorentz symmetry breaking. For our purpose, $J_{\nu}\left(y\right)=q\eta_{\nu}^{\ 0}\delta^{3}\left({\bf{y}}\right)$.

Using the Eqs. (\ref{PropM}), (\ref{propEM2}), (\ref{defAs}), (\ref{defalv}) and proceeding as before, we obtain
\begin{eqnarray}
\label{A0charge}
A^{0}\left(\mid{\bf{x}}\mid\right)&=&\frac{q}{4\pi}\left[\frac{1}{\mid{\bf{x}}\mid}+\frac{3}{2}\mid{\bf{x}}\mid\left({\bf {k}}_{AF}\right)^{2}-\frac{1}{2}\frac{\left[\left({\bf {k}}_{AF}\right)\cdot{\bf{x}}\right]^{2}}{\mid{\bf{x}}\mid}\right] \ , \\
\label{AEcharge}
{\bf{A}}\left({\bf{x}}\right)&=&\frac{q}{4\pi}\left[\frac{\left[{\bf{x}}\times\left({\bf {k}}_{AF}\right)\right]}{\mid{\bf{x}}\mid}-\frac{1}{2}\left(k_{AF}\right)^{0}\left(\frac{\left[\left({\bf {k}}_{AF}\right)\cdot{\bf{x}}\right]}{\mid{\bf{x}}\mid}{\bf{x}}-3\mid{\bf{x}}\mid\left({\bf {k}}_{AF}\right)\right)\right] \ .
\end{eqnarray}

We notice that the above potentials are static and give  an electric field as well as a magnetic field produced by a point-like charge.     

The electric field can be obtained as follows
\begin{eqnarray}
\label{elecfkch}
{\bf{E}}\left({\bf{x}}\right)&=&-{\bf{\nabla}}_{\bf{x}}\left(A^{0}\left(\mid{\bf{x}}\mid\right)\right)\nonumber\\
&=&\frac{q}{4\pi}\left[\left(\frac{1}{{\bf{x}}^{2}}-\frac{3}{2}\left({\bf {k}}_{AF}\right)^{2}-\frac{1}{2}\frac{\left[\left({\bf {k}}_{AF}\right)\cdot{\bf{x}}\right]^{2}}{{\bf{x}}^{2}}\right){\hat{x}}+\frac{\left[\left({\bf {k}}_{AF}\right)\cdot{\bf{x}}\right]}{\mid{\bf{x}}\mid}\left({\bf {k}}_{AF}\right)\right] \ ,
\end{eqnarray}
where ${\hat{x}}$ is an unit vector pointing in the direction of the vector ${\bf {x}}$.

The electric field (\ref{elecfkch}) is composed by the standard result obtained in Maxwell electrodynamics plus a correction, which we calculate up to the second order in the background vector.  The modulus of the resulting electric field reads
\begin{eqnarray}
\label{elecfkchm}
\mid{\bf{E}}\left({\bf{x}}\right)\mid =\frac{\mid q\mid}{4\pi{\bf{x}}^{2}}-\frac{\mid q\mid}{8\pi}\left({\bf {k}}_{AF}\right)^{2}\left[3-\cos^{2}\left(\phi\right)\right] \ ,
\end{eqnarray}
where $\phi$ is the angle between the vectors ${\bf x}$ and $\left({\bf {k}}_{AF}\right)$, with $0\leq\phi\leq 2\pi$.

The Lorentz violation correction to the electric field in Eq. (\ref{elecfkchm}) does not depend on the distance $\mid{\bf{x}}\mid$, but it depends on the orientation of the background vector with respect to the vector ${\bf{x}}$, which is given by the angle $\phi$. 

In the hydrogen atom, the electric field produced by the electron in the proton position (atomic electric field) has an experimental value $\sim 5.142 \times 10^{11}$N/C with a standard experimental uncertainty given by $\sim 78$N/C \cite{atomicelec}. Considering that the Lorentz violation is fully contained in this experimental uncertainty, from the second term on the right hand side of Eq. (\ref{elecfkchm}) we can make an overestimation in order of magnitude for the background vector, namely $\mid\left({\bf {k}}_{AF}\right)\mid\sim 10^{-11}$GeV, which is a very small quantity. 

Considering an atomic system and $\mid\left({\bf {k}}_{AF}\right)\mid\sim 10^{-42}$GeV, the Lorentz violation correction to the electric field in order of magnitude is $\mid\Delta{\bf{E}}\mid\sim 10^{-62}$N/C, which cannot be measured nowadays. 

The magnetic field can be calculated by using the Eq. (\ref{AEcharge}),
\begin{eqnarray}
\label{magfchar}
{\Delta\bf{B}}\left({\bf{x}}\right)={\bf{\nabla}}_{\bf{x}}\times {\bf{A}}\left({\bf{x}}\right)=-\frac{q}{4\pi\mid{\bf{x}}\mid}\left[\frac{\left[\left({\bf {k}}_{AF}\right)\cdot{\bf{x}}\right]}{\mid{\bf{x}}\mid}{\hat{x}}+\left({\bf {k}}_{AF}\right)-2\left(k_{AF}\right)^{0}\left[{\bf{x}}\times\left({\bf {k}}_{AF}\right)\right]\right] \ .
\end{eqnarray}

\begin{figure}[!h]
\centering \includegraphics[scale=0.30]{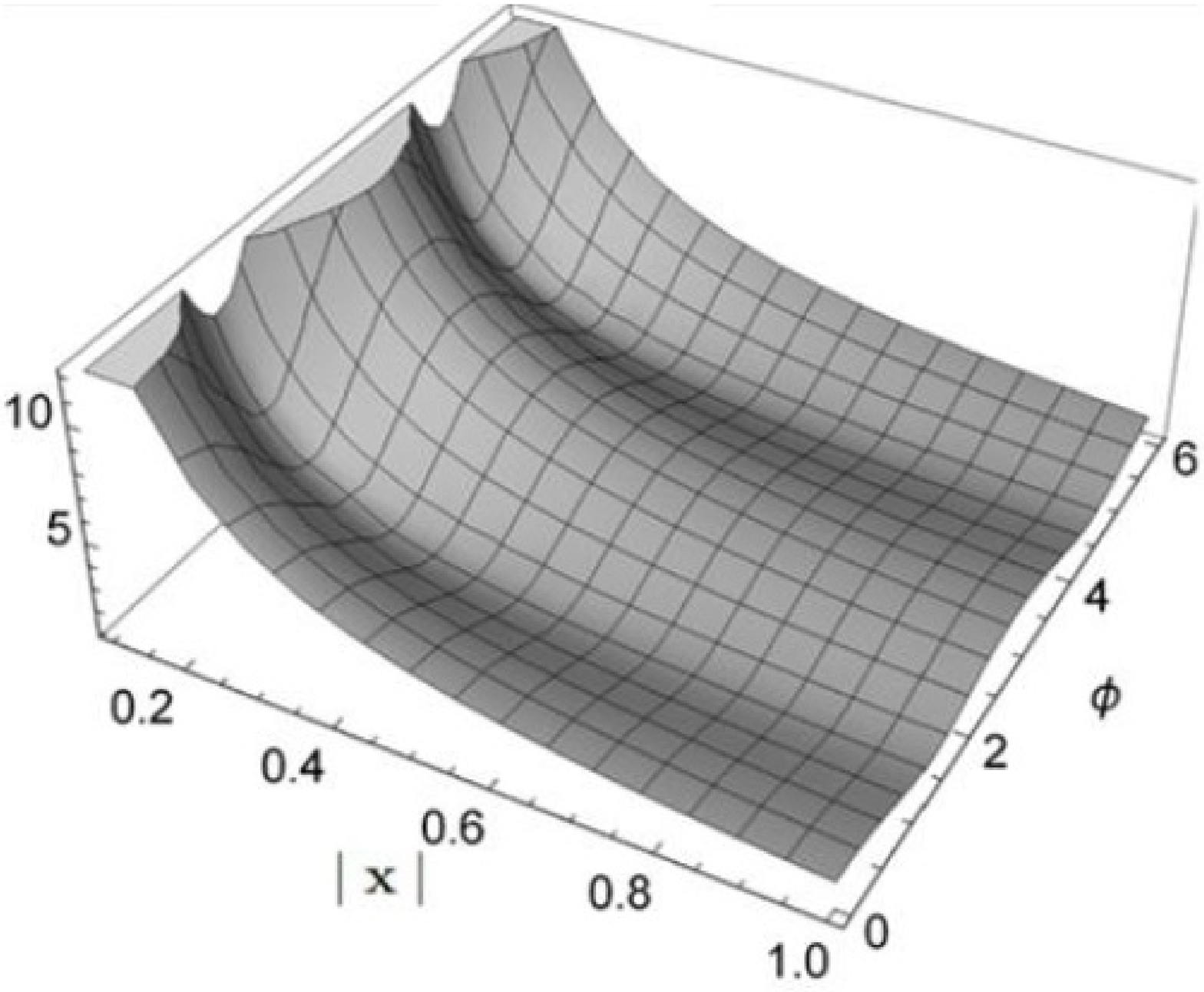} \caption{Function $f\left(\mid{\bf{x}}\mid ,\phi\right)$, appearing in (\ref{fphimag})}
\label{magch}
\end{figure}

We notice that the magnetic field (\ref{magfchar}) is an exclusive effect due to the Lorentz violation, since in Maxwell electrodynamics a static point-like charge does not produce magnetic field. This effect exhibits contributions of both first and second order in the background vector. For the case where $\left(k_{AF}\right)^{\mu}=\left(0,\left({\bf {k}}_{AF}\right)\right)$
the magnetic field becomes an effect of the first order in the background vector and its modulus reads
\begin{eqnarray}
\label{magfmchar}
\mid{\Delta\bf{B}}\left({\bf{x}}\right)\mid =\frac{\mid q\mid}{4\pi}\mid\left({\bf {k}}_{AF}\right)\mid f\left(\mid{\bf{x}}\mid ,\phi\right),
\end{eqnarray}
where we defined the function
\begin{eqnarray}
\label{fphimag}
f\left(\mid{\bf{x}}\mid ,\phi\right)=\frac{1}{\mid{\bf{x}}\mid}\left[3\cos^{2}\left(\phi\right)+1\right]^{1/2} \ .
\end{eqnarray}
The behavior of the magnetic field in (\ref{magfmchar}) is modulated by the function in (\ref{fphimag}), which depends on the orientation of $\left({\bf {k}}_{AF}\right)$ with respect to ${\bf{x}}$. In Fig. \ref{magch} we have a plot for $f\left(\mid{\bf{x}}\mid ,\phi\right)$.

Taking into account an atomic system and $\mid\left({\bf {k}}_{AF}\right)\mid\sim 10^{-42}$GeV, we can estimate the magnetic field (\ref{magfmchar}) in order of magnitude in $\mid\Delta{\bf{B}}\mid\sim 10^{-34}$T.

\subsection{Dirac strings\label{V}}

\subsubsection{Interaction energy\label{ED}}

In this subsection we search for Lorentz violating effects mediated
by the presence of Dirac strings. We start by considering a system
composed by a steady point-like charge located at position ${\bf a}$
and a Dirac string. The source for this system reads 
\begin{equation}
J_{\mu}^{DC}\left(x\right)=J_{\mu(D)}\left(x\right)+q\eta_{\ \mu}^{0}\delta^{3}({\bf x}-{\bf a})\ ,\label{Dcurrent1}
\end{equation}
with $J_{(D)}^{\mu}\left(x\right)$ standing for the source produced
by the Dirac string. The label $DC$ means that we have a Dirac string
and a point-like charge.

We choose a coordinate system where the Dirac string lies along the
$z$-axis with internal magnetic flux $\Phi$. Its source is defined
by \cite{LHCFABJHN,LHCFAB1,LHCAFFFAB1,MFXFEBFAB}, 
\begin{equation}
J_{(D)}^{\mu}(x)=i\Phi(2\pi)^{2}\int\frac{d^{4}p}{(2\pi)^{4}}\delta(p^{0})\delta(p^{3})\varepsilon_{\ \ \nu3}^{0\mu}\ p^{\nu}e^{-ipx}\ ,\label{Dircurr2}
\end{equation}
where $\Phi>0$ means an internal magnetic field pointing along $\hat{z}$,
whereas for $\Phi<0$, the internal magnetic field points in the opposite
direction.

Hereafter in this subsection, we establish that the sub-index $\perp$
means the component of a given vector perpendicular to the Dirac string.
For example, ${\bf p}_{\perp}=(p^{1},p^{2},0)$ is the momentum perpendicular
to the string. By following the same steps presented previously, from
Eqs. (\ref{Dcurrent1}), (\ref{Dircurr2}), (\ref{PropM}), (\ref{propEM2})
and (\ref{zxc1}), we obtain 
\begin{eqnarray}
E^{DC}&=&2q\Phi\Biggl[\left({k}_{AF}\right)^{3}\lim_{m\rightarrow0}\int\frac{d^{2}{\bf p}_{\perp}}{(2\pi)^{2}}\frac{e^{i{\bf p}_{\perp}\cdot{\bf a}_{\perp}}}{{\bf p}_{\perp}^{2}+m^{2}}\nonumber\\
&
&-2\left({k}_{AF}\right)^{0}\left\{ \left[{\bf {\nabla}}_{{\bf a}_{\perp}}\times\left({\bf {k}}_{AF}\right)_{\perp}\right]\cdot{\hat{z}}\right\} \lim_{m\rightarrow0}\int\frac{d^{2}{\bf p}_{\perp}}{(2\pi)^{2}}\frac{e^{i{\bf p}_{\perp}\cdot{\bf a}_{\perp}}}{\left({\bf p}_{\perp}^{2}+m^{2}\right)^{2}}\Biggr]\ ,\label{EDirac1}
\end{eqnarray}
where we introduced a mass parameter $m$, as a regulator 
in order to eliminate the divergences in a similar way to what was
done in \,\cite{LHCFABJHN,LHCFAB1,FABGFH1,LHCAFFFAB3},
and defined the operator
\begin{equation}
{\bf \nabla}_{{\bf a}_{\perp}}=\left(\frac{\partial}{\partial a^{1}},\frac{\partial}{\partial a^{2}},0\right)\ .\label{nablaperp}
\end{equation}
Using the fact that \cite{Gradshteyn}
\begin{eqnarray}
\label{IntK0}
\int\frac{d^{2}{\bf p}_{\perp}}{(2\pi)^{2}}\frac{e^{i{\bf p}_{\perp}\cdot{\bf a}_{\perp}}}{{\bf p}_{\perp}^{2}+m^{2}}&=&\frac{1}{2\pi}K_{0}\left(m\mid{\bf {a}}_{\perp}\mid\right) \ ,  \nonumber\\  
\int\frac{d^{2}{\bf p}_{\perp}}{(2\pi)^{2}}\frac{e^{i{\bf p}_{\perp}\cdot{\bf a}_{\perp}}}{\left({\bf p}_{\perp}^{2}+m^{2}\right)^{2}}&=&\frac{1}{4\pi}\frac{\mid{\bf {a}}_{\perp}\mid K_{1}\left(m\mid{\bf {a}}_{\perp}\mid\right)}{m} \ ,
\end{eqnarray}
where $K_{0}$ and $K_{1}$ stand for the K-Bessel functions \cite{Arfken}, we arrive at
\begin{eqnarray}
\label{EDDD1}
E^{DC}=\lim_{m\rightarrow0}\frac{q\Phi}{\pi}\left[\left({k}_{AF}\right)^{3}+\left({k}_{AF}\right)^{0}\left\{ \left[{\bf {a}}_{\perp}\times\left({\bf {k}}_{AF}\right)_{\perp}\right]\cdot{\hat{z}}\right\} \right]K_{0}\left(m\mid{\bf {a}}_{\perp}\mid\right) \ .
\end{eqnarray}

Considering the approximation for $K_{0}\left(m\mid{\bf {a}}_{\perp}\mid\right)$ for small arguments \cite{Arfken}, one can write
\begin{eqnarray}
\label{EDDD2}
E^{DC}&=&-\frac{q\Phi}{\pi}\left[\left({k}_{AF}\right)^{3}+\left({k}_{AF}\right)^{0}\left\{ \left[{\bf {a}}_{\perp}\times\left({\bf {k}}_{AF}\right)_{\perp}\right]\cdot{\hat{z}}\right\} \right]\nonumber\\
&
&\times\left[\ln\left(\frac{\mid{\bf {a}}_{\perp}\mid}{a_{0}}\right)+\gamma-\ln 2+\lim_{m\rightarrow0}\ln\left(ma_{0}\right)\right] \ ,
\end{eqnarray}
where $\gamma$ is the Euler constant and $a_{0}$ is an arbitrary constant length scale. Terms that not depend on the distance $\mid{\bf {a}}_{\perp}\mid$
do not contribute to the interaction force between the charge and the Dirac string, thus
they can be neglected, including this arbitrary regularizing scale. In Eq. (\ref{EDDD2}) the contribution of the second order in the background vector is divergent, so we let us  consider just the first order contribution since it becomes finite and it is the most relevant one.  Thereby, the interaction energy up to first order in $\left({k}_{AF}\right)^{\mu}$ reads  
\begin{equation}
E^{DC}=-\frac{q\Phi}{\pi}\ln\left(\frac{\mid{\bf {a}}_{\perp}\mid}{a_{0}}\right)\left({k}_{AF}\right)^{3}
\ .
\label{EDirac2}
\end{equation}

Eq. (\ref{EDirac2}) is an exclusive effect due to the Lorentz symmetry breaking, which is of the  first order in the background vector. This effect disappears if the component of $\left({k}_{AF}\right)^{\mu}$ parallel to the string vanishes. 

The energy (\ref{EDirac2}) leads to a force between the Dirac string and the charge, which falls down when $\mid{\bf {a}}_{\perp}\mid$ increases. 
The above expression leads to an spontaneous torque when we take into account the orientation of the background vector $\mid{\bf{k}}_{AF}\mid$ with respect to string. In this setup $\left(k_{AF}\right)^{3}=\mid{\bf{k}}_{AF}\mid\cos\beta$, where $0<\beta<\pi$ is the polar angle in spherical coordinates (the $z$-axis is the polar axis).

Now, we consider a system composed by two parallel Dirac strings located
a distance ${\bf {a}}_{\perp}$ apart. We choose a coordinate system
where the first string lies along the $z$ axis, with internal magnetic
flux $\Phi_{1}$ and the second one, with $\Phi_{2}$ lying along
the line ${\bf a}_{\perp}=(a^{1},a^{2},0)$. The corresponding external
source is given by 
\begin{equation}
J_{\mu}^{DD}\left({x}\right)=J_{\mu(D,1)}\left({x}\right)+J_{\mu(D,2)}\left({x}\right)\ ,\label{duasDcurrent1}
\end{equation}
where $J_{(D,1)}^{\mu}\left({x}\right)$ is found by replacing $\Phi$
by $\Phi_{1}$ in (\ref{Dircurr2}), and 
\begin{equation}
J_{(D,2)}^{\mu}\left({x}\right)=i\Phi_{2}(2\pi)^{2}\int\frac{d^{4}p}{(2\pi)^{4}}\delta(p^{0})\delta(p^{3})\varepsilon_{\ \ \nu3}^{0\mu}\ p^{\nu}e^{-ipx}e^{-i{\bf p}_{\perp}\cdot{\bf a}_{\perp}}\ ,
\end{equation}
the label $DD$ here representing the interaction between two Dirac
strings. Proceeding as in the previous cases, we can show that the
interaction energy reads 
\begin{equation}
E^{DD}=\frac{2\Phi_{1}\Phi_{2}L}{\pi}\ln\left(\frac{\mid{\bf {a}}_{\perp}\mid}{a_{0}}\right)\left\{ \left[\left({k}_{AF}\right)^{0}\right]^{2}-\left[\left({k}_{AF}\right)^{3}\right]^{2}\right\} \ ,\label{EDD1}
\end{equation}
where we identified the Dirac string length $L=\int dx^{3}$. This
effect is of the second order in the Lorentz violation parameter and
does not occur in Maxwell theory. Even if $\left({\bf {k}}_{AF}\right)_{\perp}=0$,
we have a non-vanishing effect, while for $\left({k}_{AF}\right)^{\mu}=\left(0,\left({\bf {k}}_{AF}\right)_{\perp}\right)$,
this effect disappears.

\subsubsection{Electromagnetic field outside of a Dirac string\label{EMfield}}

It is very known that in Maxwell theory the presence of a Dirac string
does not induce an electromagnetic field outside of it. Let us investigate
wether the Lorentz violation present in our theory changes this conclusion.

Let us choose a coordinate system where the Dirac string lies along
the $z$-axis, as in Eq. (\ref{Dircurr2}), and we will compute the
electromagnetic field produced by this source at an arbitrary point
${\bf {x}}_{\perp}=\left(x^{1},x^{2},0\right)$.

For the usual Maxwell electrodynamics, we substitute (\ref{Dircurr2})
and (\ref{PropM}) in (\ref{defAs}), obtaining 
\begin{equation}
A_{M(D)}^{0}\left(\mid{\bf {x}}_{\perp}\mid\right)=0\ ,\ \ \ {\bf {A}}_{M(D)}\left({\bf {x}}_{\perp}\right)=-\frac{\Phi}{2\pi{\bf {x}}_{\perp}^{2}}\left({\bf {x}}_{\perp}\times{\hat{z}}\right)\ .\label{AM}
\end{equation}
Now, with the inclusion of the Lorentz violation correction, substituting
(\ref{Dircurr2}) in (\ref{defalv}) and performing some manipulations,
we arrive at 
\begin{equation}
\Delta A_{LV(D)}^{\mu}\left({\bf {x}}_{\perp}\right)=i\Phi\int\frac{d^{2}{\bf p}_{\perp}}{(2\pi)^{2}}\ \epsilon_{\ \nu\alpha3}^{0}p^{\alpha}\Delta_{LV}^{\mu\nu}\left(p^{0}=p^{3}=0;{\bf {p}}_{\perp}\right)e^{i{\bf p}_{\perp}\cdot{\bf x}_{\perp}}\ .\label{ALV}
\end{equation}
By using the expression (\ref{propEM2}) and carrying out
the relevant integrals as before, we can show that 
\begin{equation}
\Delta A_{LV(D)}^{0}\left(\mid{\bf {x}}_{\perp}\mid\right)=-\frac{\Phi}{\pi}\ln\left(\frac{\mid{\bf {x}}_{\perp}\mid}{x_{0}}\right)\left({k}_{AF}\right)^{3}\ ,\label{A0LV}
\end{equation}
and
\begin{eqnarray}
\label{AELV}
\Delta{\bf {A}}_{LV(D)}\left(\mid{\bf {x}}_{\perp}\mid\right)=-\frac{\Phi}{\pi}\ln\left(\frac{\mid{\bf {x}}_{\perp}\mid}{x_{0}}\right)\left({k}_{AF}\right)^{0}{\hat{z}}\ ,
\end{eqnarray}
where $x_{0}$ is an arbitrary constant length scale, and we have discarded contributions that do not depend on the distance $\mid{\bf {x}}_{\perp}\mid$ since them do not contribute to the calculation of the electromagnetic field.

We notice that the potentials (\ref{A0LV}) and (\ref{AELV}) are static and exhibit only contributions of the first  order in the background vector (the contributions of the second order are divergent, therefore we are taking into account just the leading order one, which is finite). They give an electric field as well as a magnetic field outside the Dirac string.

The electric field can be computed as follows 
\begin{align}
\label{ELV}
\Delta{\bf {E}}\left({\bf {x}}_{\perp}\right)=-{\bf {\nabla}}_{{\bf x}_{\perp}}\left(\Delta A_{LV(D)}^{0}\left(\mid{\bf {x}}_{\perp}\mid\right)\right)=\frac{\Phi}{\pi\mid{\bf {x}}_{\perp}\mid}\left({k}_{AF}\right)^{3}{\hat{x}}_{\perp}
 \ ,
\end{align}
where ${\hat{x}}_{\perp}$ is an unit vector pointing in the direction of the vector ${\bf x}_{\perp}$. 

The above result is an effect due solely to the Lorentz violating background, which disappears if the component of the background vector parallel to the Dirac string is null.

The magnetic filed can be calculated in the following way 
\begin{eqnarray}
\label{BLV}
\Delta{\bf {B}}\left({\bf {x}}_{\perp}\right)={\bf {\nabla}}_{{\bf x}_{\perp}}\times\Delta{\bf {A}}_{LV(D)}\left(\mid{\bf {x}}_{\perp}\mid\right)= -\frac{\Phi}{\pi{\bf {x}}_{\perp}^{2}}\left({k}_{AF}\right)^{0}\left({\bf {x}}_{\perp}\times{\hat{z}}\right)\ .
\end{eqnarray}

It is worth mentioning that the magnetic field in Eq. (\ref{BLV}) is an exclusive effect due to the Lorentz symmetry breaking. This effect is absent if $\left(k_{AF}\right)^{\mu}=\left(0,\left({\bf{k}}_{AF}\right)\right)$.

\subsubsection{Aharonov-Bohm bound states}
\label{AB} 

The potentials (\ref{A0LV}) and (\ref{AELV}) induce a modification
in the Aharonov-Bohm bound states. In order to verify such modification,
let us consider a quantum rigid rotor \cite{Griffiths,Sakurai} composed
by a non-relativistic particle with mass $M$ and electric charge
$q$, restricted to move along a ring of radius $b$ surrounding the
Dirac string. We take a coordinate system where the ring lies on the
plane $z=0$, centered at the origin.

For simplicity, let us choose $\left(k_{AF}\right)^{\mu}=\left(0,\left(k_{AF}\right)^{3}\right)$,
where from Eqs. (\ref{fields2}), (\ref{AM}), (\ref{A0LV}) and (\ref{AELV}),
we obtain 
\begin{eqnarray}
 A_{(D)}^{0}\left(b\right) & = & -\frac{\Phi}{\pi}\ln\left(\frac{b}{b_{0}}\right)\left(k_{AF}\right)^{3}\ ,\label{AB1}\\
{\bf {A}}_{(D)}\left(b\right) & = & \frac{\Phi}{2\pi b}{\hat{\varphi}}\ ,\label{AB2}
\end{eqnarray}
where we used cylindrical coordinates, with the radial coordinate
$b=\mid{\bf x}_{\perp}\mid=\sqrt{(x^{1})^{2}+(x^{2})^{2}}$ and
with $\hat{\varphi}$ standing for the unitary vector for the azimuthal
coordinate.

The Hamiltonian for the charged particle in cylindrical coordinates
reads, 
\begin{eqnarray}
\label{AB3}
H = -\frac{1}{2Mb^{2}}\frac{d^{2}}{d\varphi^{2}}+\frac{iq\Phi}{2\pi Mb^{2}}\frac{d}{d\varphi}
+\frac{q^{2}\Phi^{2}}{8\pi^{2}Mb^{2}}-\frac{q\Phi}{\pi}\ln\left(\frac{b}{b_{0}}\right)\left(k_{AF}\right)^{3}\ .
\end{eqnarray}

The energy eigenfunctions of the Hamiltonian (\ref{AB3}) are given by 
\begin{eqnarray}
\Psi\left(\varphi\right)=Be^{in\varphi}\ \ ,\ \ n=0,\pm1,\pm2,\cdots\label{eigenfun}
\end{eqnarray}
where $B$ is a normalization constant.  The corresponding energy levels are 
\begin{eqnarray}
E_{n} & = & \frac{1}{2Mb^{2}}\left(n-\frac{q\Phi}{2\pi}\right)^{2}-\frac{q\Phi}{\pi}\ln\left(\frac{b}{b_{0}}\right)\left(k_{AF}\right)^{3} \ .\label{enerahaketr}
\end{eqnarray}

The first term on the right hand side of (\ref{enerahaketr}) is the
well known Aharonov Bohm energy \cite{Griffiths}, and the second one is a correction at first  order in the background vector. For the photon CPT-even sector of the SME the corrections to the Aharonov-Bohm bound states were studied in Ref. \cite{LHCFAB1}.

\subsection{Point-like dipoles}
\label{dipoles} 

In this subsection we consider other interesting system consisting
of two stationary point-like dipoles located at fixed points ${\bf {a}}_{1}$
and ${\bf {a}}_{2}$, respectively. The external source which describes
such a system reads \cite{FABGFH1}, 
\begin{eqnarray}
J_{\mu}^{dd}\left(x\right)=\eta_{\ \mu}^{0}U^{\beta}\partial_{\beta}\left[\delta^{3}\left({\bf {x}}-{\bf {a}}_{1}\right)\right]+\eta_{\ \mu}^{0}W^{\beta}\partial_{\beta}\left[\delta^{3}\left({\bf {x}}-{\bf {a}}_{2}\right)\right]\ ,\label{CDipole}
\end{eqnarray}
where $U^{\beta}=\left(0,{\bf {U}}\right)$ and $W^{\beta}=\left(0,{\bf {W}}\right)$
are fixed and static four vectors in the reference frame we are performing
the calculations, with ${\bf {U}}$ and ${\bf {W}}$ standing for
the dipole moments. The super index $dd$ means that we have two dipoles.

Following the same steps employed in the previous subsections, we
find for the interaction energy between the two dipoles, 
\begin{eqnarray}
\label{Eddipole}
E^{dd} & = & \frac{1}{4\pi\mid{\bf {a}}\mid^{3}}\Biggl\{\left[\left({\bf {U}}\cdot{\bf {W}}\right)-3\frac{\left({\bf {U}}\cdot{\bf {a}}\right)\left({\bf {W}}\cdot{\bf {a}}\right)}{{\bf {a}}^{2}}\right]-\frac{3{\bf {a}}^{2}}{2}\left({\bf {k}}_{AF}\right)^{2}\left[\left({\bf {U}}\cdot{\bf {W}}\right)-\frac{\left({\bf {U}}\cdot{\bf {a}}\right)\left({\bf {W}}\cdot{\bf {a}}\right)}{{\bf {a}}^{2}}\right]\nonumber \\
 &  & +{\bf {a}}^{2}\left[\left({\bf {k}}_{AF}\right)\cdot{\bf {U}}\right]\left[\left({\bf {k}}_{AF}\right)\cdot{\bf {W}}\right]-\left[\left({\bf {k}}_{AF}\right)\cdot{\bf {a}}\right]\Bigl\{\left[\left({\bf {k}}_{AF}\right)\cdot{\bf {W}}\right]\left({\bf {U}}\cdot{\bf {a}}\right)+\left[\left({\bf {k}}_{AF}\right)\cdot{\bf {U}}\right]\left({\bf {W}}\cdot{\bf {a}}\right)\Bigr\}\nonumber \\
 &  & -\frac{\left[\left({\bf {k}}_{AF}\right)\cdot{\bf {a}}\right]^{2}}{2}\left[\left({\bf {U}}\cdot{\bf {W}}\right)+9\frac{\left({\bf {U}}\cdot{\bf {a}}\right)\left({\bf {W}}\cdot{\bf {a}}\right)}{{\bf {a}}^{2}}\right]\Biggr\}\ .
\end{eqnarray}
where ${\bf {a}}={\bf {a}}_{1}-{\bf {a}}_{2}$ is the distance between
the two dipoles.

The first contribution between brackets on the right hand side of
Eq. (\ref{Eddipole}) is the well-known result obtained in standard
Maxwell theory \cite{FABGFH1}, the remaining terms are second order corrections
in the background vector. We can also analyse different particular
cases and torques depending on the orientation of the dipoles relative
to the background vector.


\section{Nonminimal CPT-odd sector\label{NM}}


In this section we consider a model which belongs to the CPT-odd sector
of the nonminimal SME. The Lagrangian density is given by \cite{MMFCPTodd}
\begin{eqnarray}
{\cal L}=-\frac{1}{4}F_{\mu\nu}F^{\mu\nu}-\frac{1}{2\gamma}\left(\partial_{\mu}A^{\mu}\right)^{2}+\frac{1}{2}\epsilon^{\mu\nu\alpha\beta}V_{\mu}A_{\nu}\Box F_{\alpha\beta}+J^{\mu}A_{\mu}\ .\label{lagEmNM}
\end{eqnarray}

We notice that the Lorentz violating term in (\ref{lagEmNM}) has mass dimension 5, and the Lorentz breaking coefficient $V^{\mu}$
has inverse of mass dimension. We can map this term into the ones
presented in reference \cite{VAKZL} for the the nonminimal SME. The
underlying Lagrangian must be ${\cal {L}}_{A}^{(5)}$ of their Tab.
III, 
\begin{eqnarray}
{\cal {L}}_{A}^{(5)}=-\frac{1}{4}k^{(5)\beta\kappa\lambda\rho\phi}F_{\kappa\lambda}\partial_{\beta}F_{\rho\phi}\ ,\label{map}
\end{eqnarray}
with the background tensor restricted to the form, 
\begin{eqnarray}
k^{(5)\beta\kappa\lambda\rho\phi}=\eta^{\beta\kappa}\epsilon^{\mu\lambda\rho\phi}V_{\mu}\ .\label{K5T}
\end{eqnarray}

In the course of this section we will use a notation similar to that
one employed in the previous section, and we will treat the background
vector perturbatively up to second order.

The propagator for the model (\ref{lagEmNM}) for $\gamma=1$ up to second order in $V^{\mu}$  is given by the expression (\ref{propEm}), with 
\begin{eqnarray}
\label{propNM}
\Delta_{LV}^{\mu\nu}\left(p\right) & = & -\frac{1}{p^{2}}\Biggl[\eta^{\mu\nu}\left[4\left(V\cdot p\right)^{2}-4V^{2}p^{2}\right]-2i\epsilon^{\mu\nu\alpha\beta}V_{\alpha}p_{\beta}+4V^{2}p^{\mu}p^{\nu}\nonumber \\
 &  & +4p^{2}V^{\mu}V^{\nu}-4\left(V\cdot p\right)\left(V^{\mu}p^{\nu}+V^{\nu}p^{\mu}\right)\Biggr]\ ,
\end{eqnarray}
which is in perfect agreement with Eq. (20a) of the Ref. \cite{MMFCPTodd}. In order to check this information, we must replace in the expression (20a), $\xi=-1, D^{\mu}\rightarrow V^{\mu}$ and then we carry out an expansion up to second order in the Lorentz violation parameter.

The interaction energy between different pairs of electromagnetic
external sources for the theory (\ref{lagEmNM}), can also be calculated
from the Eq. (\ref{zxc1}).

\subsection{Point-like charges}
\label{NMcharges} 

\subsubsection{Interaction energy and torque\label{ETorqueNM}}

In this subsection we consider the interaction energy between two
charges mediated by the model (\ref{lagEmNM}). Substituting (\ref{corre1Em}),
(\ref{PropM}), (\ref{propNM}) in (\ref{zxc1}) and following similar
steps employed in previous section, we obtain 
\begin{eqnarray}
\label{ECCNMMM}
E^{CC}&=&q_{1}q_{2}\left[\int\frac{d^{3}{\bf p}}{(2\pi)^{3}}\frac{e^{i{\bf p}\cdot{\bf a}}}{{\bf {p}}^{2}}-4\left({\bf{V}}\cdot{\bf {\nabla}}_{{\bf {a}}}\right)^{2}\int\frac{d^{3}{\bf p}}{(2\pi)^{3}}\frac{e^{i{\bf p}\cdot{\bf a}}}{{\bf {p}}^{2}}-4{\bf {V}}^{2}\int\frac{d^{3}{\bf p}}{(2\pi)^{3}}e^{i{\bf p}\cdot{\bf a}}\right]\\
\label{ECCNM}
&=&\frac{q_{1}q_{2}}{4\pi\mid{\bf {a}}\mid}\left[1-\frac{4}{{\bf {a}}^{2}}\left(3\frac{\left({\bf {V}}\cdot{\bf {a}}\right)^{2}}{{\bf {a}}^{2}}-{{\bf {V}}^{2}}\right)\right]\ ,
\end{eqnarray}
where the last term inside the brackets of Eq. (\ref{ECCNMMM}) is the Dirac delta function $\delta^{3}({{\bf{a}}})$ and provided that ${\bf{a}}$ is nonzero, this term vanishes.

We notice that Eq. (\ref{ECCNM}) is a perturbative result up to second order in
the Lorentz breaking parameter. The first term between brackets on
the right hand side stands for the Coulomb interaction and the remaining
ones are Lorentz violation corrections. 

\begin{figure}[!h]
\centering \includegraphics[scale=0.35]{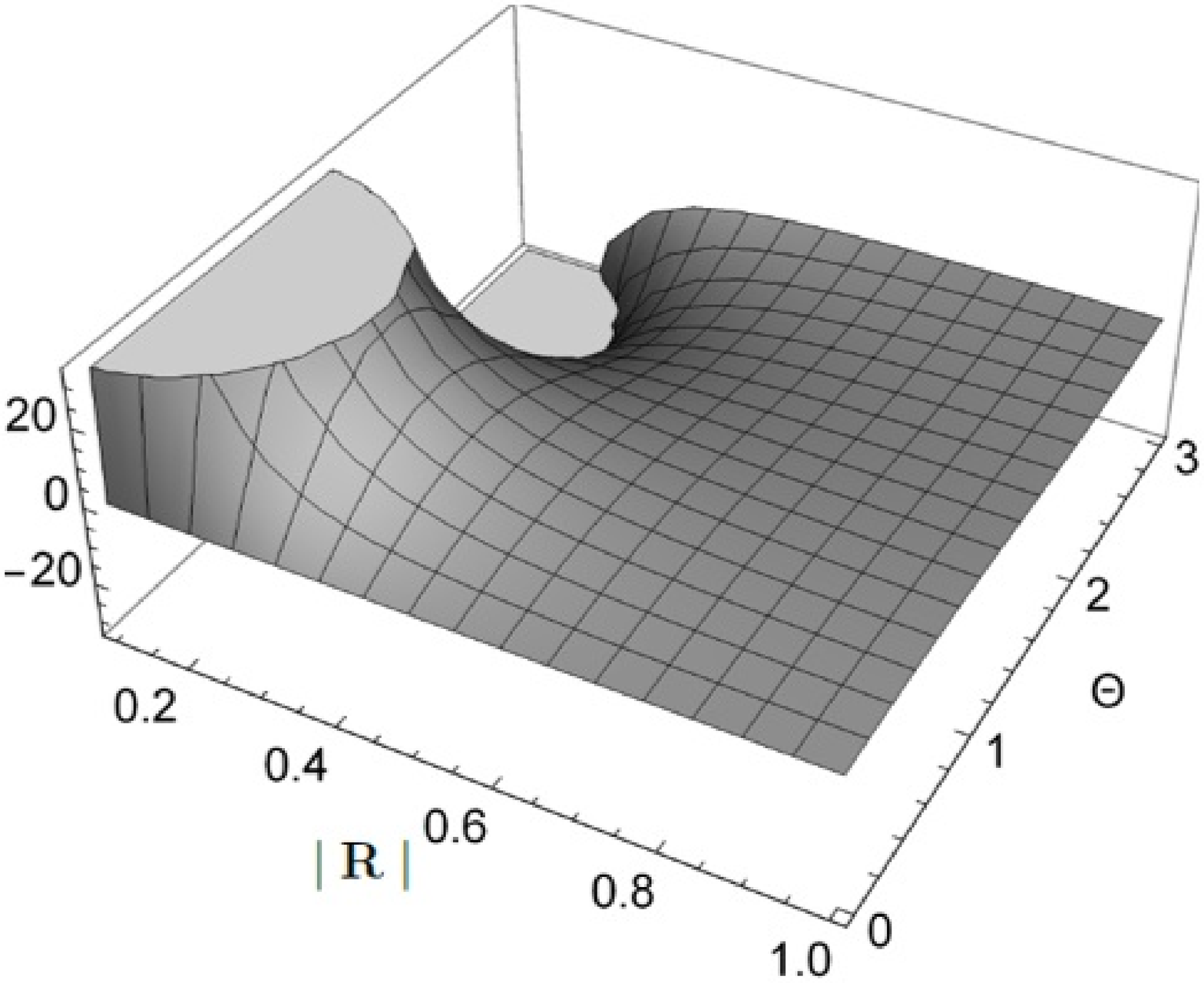} \caption{Torque (\ref{TorqueNM}) multiplied by $\frac{\pi}{3q^{2}{\bf {v}}^{2}}$.}
\label{grafico9}
\end{figure}

Considering an usual dipole
with $q_{1}=-q_{2}=q$, placed a distance $\mid{\bf {R}}\mid$ apart,
and defining by $0\leq\Theta\leq\pi$ the angle between ${\bf {V}}$
and ${\bf {R}}$, the energy (\ref{ECCNM}) leads to a spontaneous
torque on this dipole, as follows 
\begin{eqnarray}
\tau_{dipole}^{CC}=\frac{3q^{2}{\bf {V}}^{2}}{\pi\mid{\bf {R}}\mid^{3}}\sin\left(2\Theta\right)\ ,\label{TorqueNM}
\end{eqnarray}
which vanishes for $\Theta=0,\pi,\pi/2$ and ${\bf {V}}=0$. In Fig.
\ref{grafico9}, we have a plot of (\ref{TorqueNM}) multiplied by
$\frac{\pi}{3q^{2}{\bf {v}}^{2}}$.

In Tab. D17 of reference \cite{D17CPTodd}, we can find upper bounds
imposed on some Lorentz violating parameters models involving terms
with dimension 5 of mass. In particular in \cite{AltschulRC} $V^{\mu}$
was overestimated through radiative corrections in the range $\left(10^{-25}-10^{-31}\right)$GeV$^{-1}$
(see also \cite{D17CPTodd}). Using these estimates for $\mid{\bf {V}\mid}$,
and considering an atomic system, where $q\sim1.60217\times10^{-19}$C
and $\mid{\bf {R}}\mid\sim10^{-10}$m, we have $\tau_{dipole}^{CC}\sim\left(10^{-78}-10^{-90}\right)$Nm,
which is really out of any experimental reach nowadays. However, we
can find stricter upper bounds for operators with dimension 5 through
astrophysical birefringence measurements, which can reach $\sim10^{-35}$
GeV$^{-1}$ \cite{D17CPTodd,VAKMMAB}, where for this estimate, $\tau_{dipole}^{CC}\sim10^{-98}$Nm.

\subsubsection{Electromagnetic field configuration}

\label{EMfieldcharge} 

Now, let us obtain the electromagnetic field configuration generated
by a steady point-like charge located at the origin. We will evaluate
the field produced by this charge at an arbitrary non-null point ${\bf {r}}=\left(r^{1},r^{2},r^{3}\right)$.

Using Eqs. (\ref{fields2}), (\ref{defAs}), (\ref{defalv}) (\ref{PropM}), (\ref{propNM}) and considering
the external source of a stationary charge placed at origin, we find
the potentials, 
\begin{eqnarray}
A^{0}\left(\mid{\bf {r}}\mid\right)=\frac{q}{4\pi}\left[\frac{1}{\mid{\bf {r}}\mid}-\frac{4}{\mid{\bf {r}}\mid^{3}}\left(3\frac{\left({\bf {V}}\cdot{\bf {r}}\right)^{2}}{{\bf {r}}^{2}}-{\bf {V}}^{2}\right)\right]\ ,\label{A0CNM}
\end{eqnarray}
\begin{eqnarray}
{\bf {A}}\left({\bf {r}}\right)=\frac{q}{2\pi\mid{\bf {r}}\mid^{3}}\left[\left({\bf {r}}\times{\bf {V}}\right)-2V^{0}\left(3\frac{\left({\bf {V}}\cdot{\bf {r}}\right)}{\mid{\bf {r}}\mid}{\hat{r}}-{\bf {V}}\right)\right]\ ,\label{AECNM}
\end{eqnarray}
where ${\hat{r}}$ is the unit vector pointing on the direction of the vector ${\bf {r}}$. Since the vector ${\bf{r}}$ is not vanishing, we neglected contributions proportional to $\delta^{3}({{\bf{r}}})$.

The expression (\ref{A0CNM}) leads to an electric field, as follows
\begin{eqnarray}
{\bf {E}}\left({\bf {r}}\right)=\frac{q}{4\pi{\bf {r}}^{2}}\Biggl\{\left[1-\frac{12}{{\bf {r}}^{2}}\left(5\frac{\left({\bf {V}}\cdot{\bf {r}}\right)^{2}}{{\bf {r}}^{2}}-{\bf {V}}^{2}\right)\right]{\hat{r}}+24\frac{\left({\bf {V}}\cdot{\bf {r}}\right)}{\mid{\bf {r}}\mid^{3}}{\bf {V}}
\Biggr\}\ ,\label{ENM}
\end{eqnarray}
whose modulus is given by, 
\begin{eqnarray}
\mid{\bf {E}}\left({\bf {r}}\right)\mid=\frac{\mid q\mid}{4\pi{\bf {r}}^{2}}+\frac{3\mid q\mid{\bf {V}}^{2}}{\pi}F\left(\mid{\bf {r}}\mid,\chi\right)\ ,\label{MENM}
\end{eqnarray}
where we defined the function, 
\begin{eqnarray}
F\left(\mid{\bf {r}}\mid,\chi\right)=\frac{1}{\mid{\bf {r}}\mid^{4}}\left[1-3\cos^{2}\left(\chi\right)\right]\ ,\label{FNM}
\end{eqnarray}
with $0\leq\chi<2\pi$ standing for the angle between ${\bf {r}}$
and ${\bf {V}}$.

\begin{figure}[!h]
\centering \includegraphics[scale=0.35]{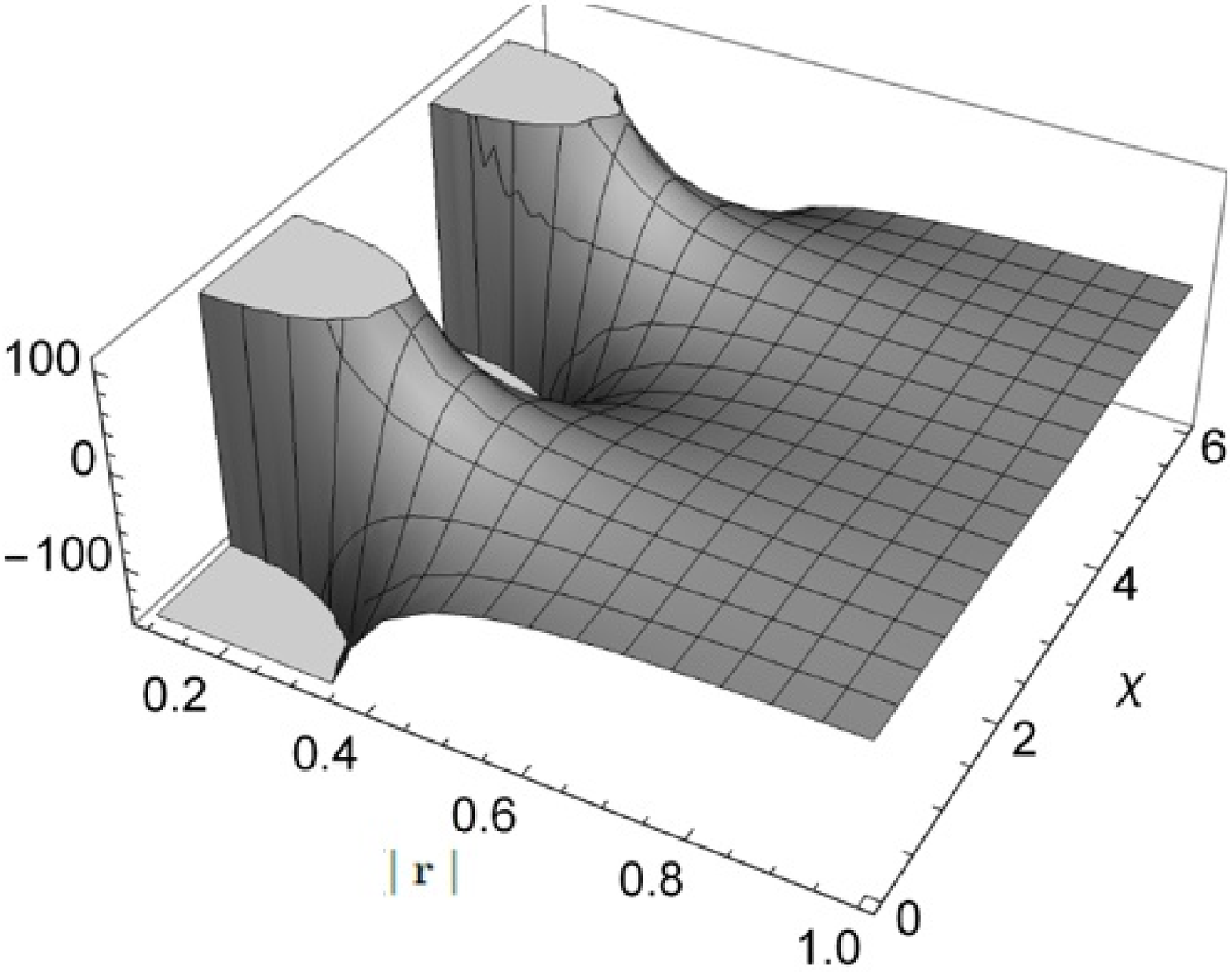} \caption{Function $F\left(\mid{\bf {r}}\mid,\chi\right)$, appearing
in (\ref{FNM}).}
\label{grafico11}
\end{figure}

The first term on the right hand side of Eq. (\ref{MENM}) is the
electric field found in standard Maxwell electrodynamics, The second
one is a correction up to second order in the Lorentz breaking
parameter whose behavior is given by the function in (\ref{FNM}).
In Fig. \ref{grafico11} is showed a plot for the function in (\ref{FNM}).

Considering an atomic system, we have that the Lorentz violation correction
for the electric field, in order of magnitude, is $\mid\Delta{\bf {E}}\mid\sim\left(10^{-49}-10^{-60}\right)$N/C
(radiative corrections) and $\mid\Delta{\bf {E}}\mid\sim10^{-69}$N/C (astrophysical birefringence).

Proceeding similarly to what we did in the subsection \ref{III}, we can overestimate the background vector assuming that the Lorentz violation is contained in the experimental uncertainty value of the atomic electric field, namely $\sim 78$N/C \cite{atomicelec}.  Thus, from the second term on the right hand side of Eq. (\ref{MENM}), we have $\mid{\bf {V}}\mid\sim 0.68 $GeV$^{-1}$.

From Eq. (\ref{AECNM}), we can show that the magnetic field produced
by the charge reads 
\begin{eqnarray}
\Delta{\bf {B}}\left({\bf {r}}\right)=-\frac{q}{2\pi\mid{\bf {r}}\mid^{3}}\left[3\frac{\left({\bf {V}}\cdot{\bf {r}}\right)}{\mid{\bf {r}}\mid}{\hat{r}}-{\bf {V}}\right]\ .\label{BNM}
\end{eqnarray}

\begin{figure}[!h]
\centering \includegraphics[scale=0.30]{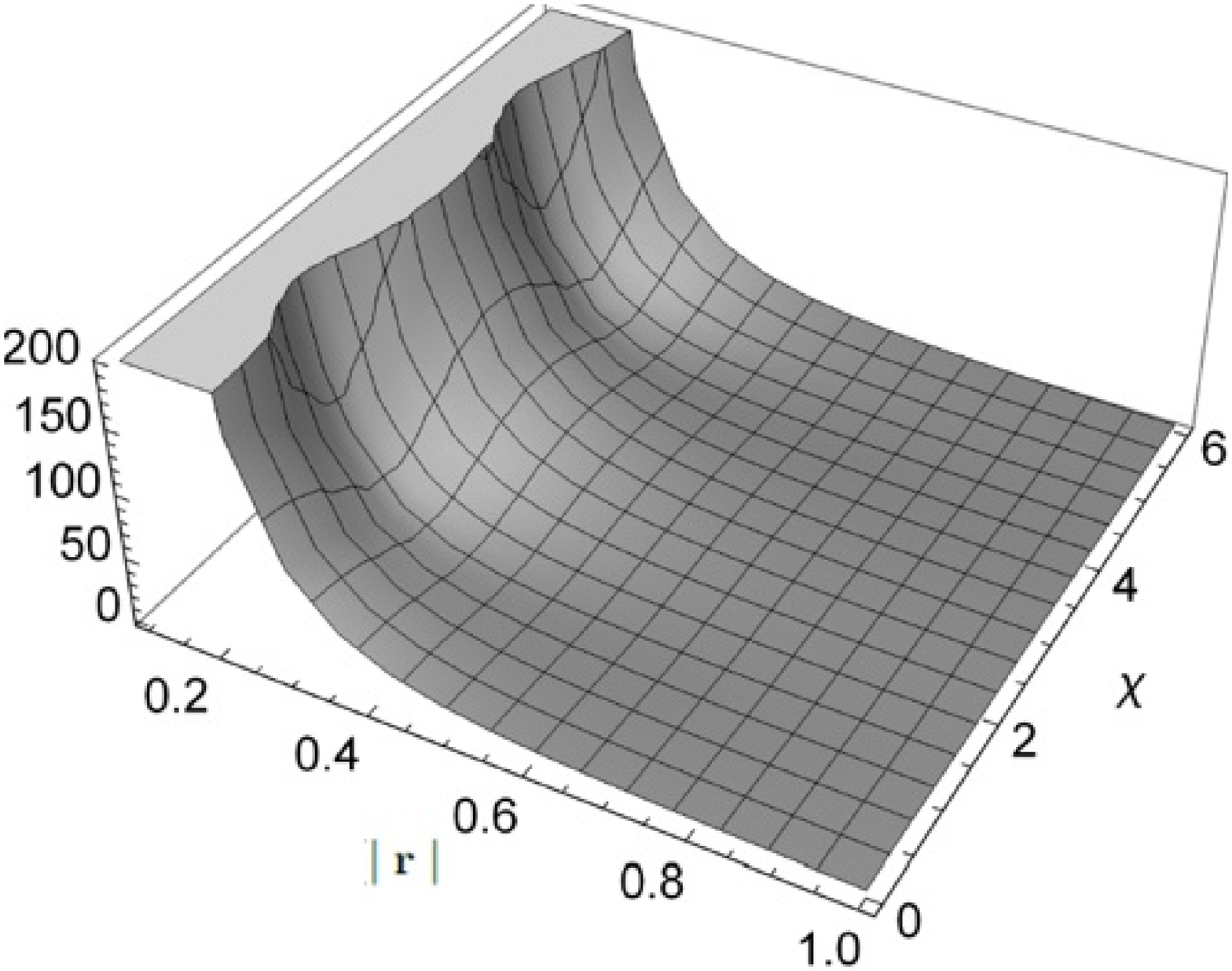} \caption{Function $G\left(\mid{\bf {r}}\mid,\chi\right)$, appearing
in (\ref{GNM}).}
\label{grafico12}
\end{figure}

It is noteworthy that a stationary point-like
charge produces a magnetic field, which is a new effect that has no
counterpart in standard Maxwell theory. We notice that this effect
is of the first order in $V^{\mu}$, which vanishes if ${\bf {V}}=0$.
The magnetic field modulus is given by, 
\begin{eqnarray}
\mid\Delta{\bf {B}}\left({\bf {r}}\right)\mid=\frac{\mid q\mid\mid{\bf {V}}\mid}{2\pi}G\left(\mid{\bf {r}}\mid,\chi\right)\ ,\label{MBNM}
\end{eqnarray}
where 
\begin{eqnarray}
G\left(\mid{\bf {r}}\mid,\chi\right)=\frac{1}{\mid{\bf {r}}\mid^{3}}\left[3\cos^{2}\left(\chi\right)+1\right]^{1/2}\ ,\label{GNM}
\end{eqnarray}
whose plot is showed in Fig. \ref{grafico12}.

For a atomic system, we have in order of magnitude, $\mid\Delta{\bf {B}}\mid\sim\left(10^{-28}-10^{-34}\right)$T
(radiative corrections) and $\mid\Delta{\bf {B}}\mid\sim10^{-38}$T
(astrophysical birefringence).

\subsection{The steady current line}
\label{SCLNM} 

\subsubsection{Interaction energy and torque\label{ETorqueSCL}}

We start this subsection considering the interaction energy between a steady
line current and a point-like stationary charge, a situation where
no force or torque appear in the standard Maxwell case, but they can
be present in a LV scenario, as discussed in\,\cite{LHCFABJHN,LHCAFFFAB1,LHCFAB1}.
Let us choose a coordinate system where the current line flows parallel
to the $z$-axis, along the straight line which crosses the $xy$
plane at ${\bf R}=(R^{1},R^{2},0)$. The electric charge is located
at an arbitrary position ${\bf {s}}$. This setup is defined by the
external source, 
\begin{equation}
J_{\mu}^{SC}\left(x\right)=I\eta_{\ \mu}^{3}\delta^{2}\left({\bf x}_{\perp}-{\bf R}\right)+q\eta_{\ \mu}^{0}\delta^{3}\left({\bf x}-{\bf s}\right)\ ,\label{corre3Em}
\end{equation}
where ${\bf x}_{\perp}=(x^{1},x^{2},0)$ stands for the position vector
perpendicular to the steady current line. The parameters $I$ and
$q$ are the current intensity and the electric charge, respectively.
The label $SC$ means that we are looking into the interaction involving
a current line and a point-like charge.

Proceeding as before, from Eqs. (\ref{corre3Em}), (\ref{PropM}) and (\ref{zxc1}) we
obtain the following interaction energy,
\begin{eqnarray}
E^{SC}=\frac{Iq}{2}\int\frac{d^{2}{\bf p}_{\perp}}{(2\pi)^{2}}\left[\Delta_{LV}^{30}\left(p^{0}=p^{3}=0;{\bf {p}}_{\perp}\right)e^{i{\bf p}_{\perp}\cdot{\bf a}_{\perp}}+\Delta_{LV}^{03}\left(p^{0}=p^{3}=0;{\bf {p}}_{\perp}\right)e^{-i{\bf p}_{\perp}\cdot{\bf a}_{\perp}}\right]\ ,\label{Ener8EM}
\end{eqnarray}
with ${\bf p}_{\perp}=(p^{1},p^{2},0)$, and the distance between
the charge and the current line defined by ${\bf a}_{\perp}={\bf R}-{\bf s}_{\perp}=(R^{1}-s^{1},R^{2}-s^{2},0)$.
Making use of Eq. (\ref{propNM}), we obtain
\begin{eqnarray}
\label{escki}
E^{SC}&=&\lim_{m\rightarrow0}\frac{Iq}{\pi}\left[\left({\bf {a}}_{\perp}\times{\bf {V}}_{\perp}\right)\cdot{\hat{z}}\right]\left(\frac{mK_{1}\left(m\mid{\bf {a}}_{\perp}\mid\right)}{\mid{\bf a}_{\perp}\mid}\right)\nonumber\\
\label{ESCNM}
&=&\frac{Iq}{\pi{\bf {a}}_{\perp}^{2}}\left[\left({\bf {a}}_{\perp}\times{\bf {V}}_{\perp}\right)\cdot{\hat{z}}\right] \ .
\end{eqnarray}

This interaction energy is an effect due solely to the Lorentz symmetry breaking, which is of the first order in the background vector. If ${\bf {V}}_{\perp}$ is null or parallel to ${\bf {a}}_{\perp}$, this effect disappears.

It is important to highlight that up to second order in the background vector the interaction energy between a steady current line and a point-like charge is divergent for the model (\ref{lagEm}). Following the similar steps employed previously, this fact can be checked in the following way
\begin{eqnarray}
\label{ESCMINIMO}
E^{SC} &=&-2Iq\Bigl[\left\{ \left[{\bf {\nabla}}_{{\bf a}_{\perp}}\times\left({\bf {k}}_{AF}\right)_{\perp}\right]\cdot{\hat{z}}\right\} +2\left({k}_{AF}\right)^{0}\left({k}_{AF}\right)^{3}\Bigr]\lim_{m\rightarrow0}\int\frac{d^{2}{\bf p}_{\perp}}{(2\pi)^{2}}\frac{e^{i{\bf p}_{\perp}\cdot{\bf a}_{\perp}}}{\left({\bf p}_{\perp}^{2}+m^{2}\right)^{2}} \nonumber\\
&=&-\frac{Iq}{2\pi}\Biggl[-\{\left[{\bf {a}}_{\perp}\times\left({\bf {k}}_{AF}\right)_{\perp}\right]\cdot{\hat{z}}\}\lim_{m\rightarrow0}K_{0}\left(m\mid{\bf {a}}_{\perp}\mid\right)\nonumber\\
&
&+2\left({k}_{AF}\right)^{0}\left({k}_{AF}\right)^{3}\lim_{m\rightarrow0}\frac{\mid{\bf {a}}_{\perp}\mid K_{1}\left(m\mid{\bf {a}}_{\perp}\mid\right)}{m}\Biggr] \nonumber\\
&=&-\frac{Iq}{2\pi}\Biggl[\{\left[{\bf {a}}_{\perp}\times\left({\bf {k}}_{AF}\right)_{\perp}\right]\cdot{\hat{z}}\}\left[\ln\left(\frac{\mid{\bf {a}}_{\perp}\mid}{a_{0}}\right)+\gamma-\ln 2+\lim_{m\rightarrow0}\ln\left(ma_{0}\right)\right]\nonumber\\
&
&+2\left({k}_{AF}\right)^{0}\left({k}_{AF}\right)^{3}\lim_{m\rightarrow0}\frac{1}{m^{2}}\Biggr] \ .
\end{eqnarray}

From Eq. (\ref{ESCMINIMO}) we can notice that the contribution of the second order in the background vector can be neglected, sice it does not depend on the distance, however the first order contribution is infrared divergent. In summary, the electrostatic energy between a point charge and a current line is better behaved in the infrared for the nonminimal model \eqref{lagEmNM} than for the minimal model \eqref{lagEm}, which is interesting because higher derivative theories are known for having a better \emph{ultraviolet} behavior in general. As for the infrared, more derivatives in the Lagrangian means more momentum powers in the denominator of the propagator, potentially leading to worse infrared behavior. This is not what happens in the models studied in this work, and the reason is they have different dimensionful expansion parameters: $k_{AF}$ with dimension of mass for \eqref{lagEm}, and $V$ with dimension of inverse mass for \eqref{lagEmNM}.

Defining by $\Lambda\in\left[0,2\pi\right)$ the angle between ${\bf {V}}_{\perp}$
and ${\bf {a}}_{\perp}$, and fixing the position vector ${\bf {a}}_{\perp}$,
one can show that the energy (\ref{ESCNM}) leads us to a spontaneous torque in the whole system, as follows 
\begin{eqnarray}
\tau^{SC}=-\frac{Iq}{\pi\mid{\bf {a}}_{\perp}\mid}\mid{\bf {V}}_{\perp}\mid\cos\left(\Lambda\right)\ .\label{torqueSCNM}
\end{eqnarray}

\begin{figure}[!h]
\centering \includegraphics[scale=0.35]{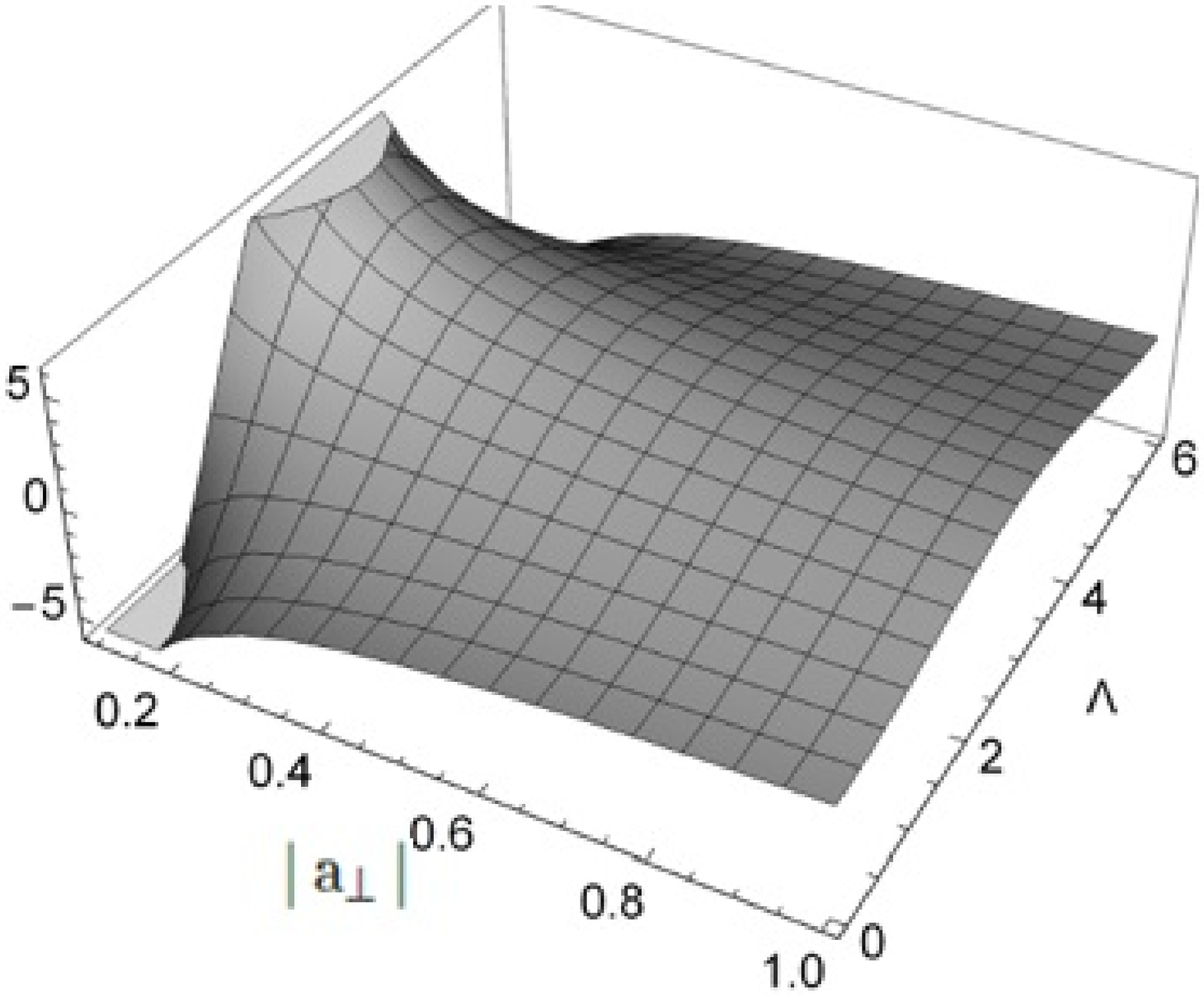} \caption{Torque (\ref{torqueSCNM}) multiplied by $\frac{\pi}{Iq\mid{\bf {V}}_{\perp}\mid}$.}
\label{grafico13}
\end{figure}

As expected, this torque vanishes when Lorentz symmetry is restored, as well as for the special configuration with $\Lambda=\pi/2$. In Fig. \ref{grafico13}, we have a plot for the torque (\ref{torqueSCNM}) multiplied by $\frac{\pi}{Iq\mid{\bf {V}}_{\perp}\mid}$.

Being first order instead of second order in the LV parameter, the
presence of this torque should be closer to the possibility of experimental
investigations. Indeed, assuming again an atomic system  as well as the fact that the highest electric currents obtained in laboratory are of magnitude
$\sim10^{5}$A, we obtain  $\tau^{SC}\sim\left(10^{-43}-10^{-49}\right)$Nm (radiative corrections) and $\tau^{SC}\sim 10^{-53}$Nm (astrophysical birefringence). We can also search for Lorentz violating signals in currents
jets produced on galaxies, in this case the current can reach the
magnitude of $\sim10^{18}$A \cite{PPRVEGLL}, and we find $\tau^{SC}\sim\left(10^{-30}-10^{-36}\right)$Nm
(radiative corrections) and $\tau^{SC}\sim10^{-40}$Nm (astrophysical birefringence). In any case, these torques are still beyond the experimental possibilities in the present and near-future.

\subsubsection{Electromagnetic field outside of a steady current line\label{EMfieldSCL}}

Now, let us calculate the electromagnetic field produced by a static current line outside of it. For this task, we choose a coordinate system where the current line lies along the $z$-axis, and we evaluated the electromagnetic filed in an arbitrary nonzero point ${\bf{r}}_{\perp}=\left(r^{1},r^{2},0\right)$. The corresponding external source reads
\begin{eqnarray}
\label{currlorigin}
J_{\nu}\left(y\right)=I\eta_{\ \nu}^{3}\delta^{2}\left({\bf y}_{\perp}\right) \ .
\end{eqnarray}

Proceeding as before, we obtain  
\begin{eqnarray}
\label{potclmu}
A^{\mu}\left({\bf{r}}_{\perp}\right)=I\int\frac{d^{2}{\bf p}_{\perp}}{(2\pi)^{2}}\left[\Delta^{\mu 3}\left(p^{0}=p^{3}=0;{\bf {p}}_{\perp}\right)+\Delta_{LV}^{\mu 3}\left(p^{0}=p^{3}=0;{\bf {p}}_{\perp}\right)\right]e^{i{\bf p}_{\perp}\cdot{\bf r}_{\perp}} \ ,
\end{eqnarray}
Using the Eqs. (\ref{PropM}) and (\ref{propNM}), we obtain the potentials up to second order in the background vector,
\begin{eqnarray}
\label{A0currline}
A^{0}\left(\mid{\bf{r}}_{\perp}\mid\right)=-\frac{I}{\pi{\bf{r}}_{\perp}^{2}}\left[\left({\bf {r}}_{\perp}\times{\bf {V}}_{\perp}\right)\cdot{\hat{z}}\right] \ ,
\end{eqnarray}
and
\begin{eqnarray}
\label{AVcurreline}
{\bf{A}}\left({\bf{r}}_{\perp}\right)&=&\frac{I}{2\pi}\ln\left(\frac{\mid{\bf{r}}_{\perp}\mid}{r_{0}}\right){\hat{z}}+\frac{I}{\pi{\bf{r}}_{\perp}^{2}}\Biggl[V^{0}\left({\bf{r}}_{\perp}\times{\hat{z}}\right)-2V^{3}\left(2\frac{\left({\bf {V}}_{\perp}\cdot{\bf {r}}_{\perp}\right)}{{\bf{r}}_{\perp}^{2}}{\bf{r}}_{\perp}-{\bf{V}}_{\perp}\right)\nonumber\\
&
&+2\left(2\frac{\left({\bf {V}}_{\perp}\cdot{\bf {r}}_{\perp}\right)^{2}}{{\bf{r}}_{\perp}^{2}}-{\bf{V}}_{\perp}^{2}\right){\hat{z}}\Biggr] \ .
\end{eqnarray}

From Eq. (\ref{A0currline}), we can see that in Maxwell electrodynamics a steady current line does not produces an electric filed outside of it, this is not the case in   the presence of the background vector. The first term on the right hand side of Eq. (\ref{AVcurreline}) stands for vector potential in standard Maxwell theory, the additional contributions are corrections imposed by the Lorentz symmetry breaking, which are of the second order in the background vector.

The Eq. (\ref{A0currline}) leads to an electric field outside of the current line given by
\begin{eqnarray}
\label{eleccline}
{\Delta\bf{E}}\left({\bf{r}}_{\perp}\right)=-\frac{I}{\pi{\bf{r}}_{\perp}^{2}}\left[2\frac{\left[\left({\bf {r}}_{\perp}\times{\bf {V}}_{\perp}\right)\cdot{\hat{z}}\right]}{\mid{\bf{r}}_{\perp}\mid}{\hat{r}}_{\perp}-\left({\bf{V}}_{\perp}\times{\hat{z}}\right)\right] \ ,
\end{eqnarray}
where ${\hat{r}}_{\perp}$ is the unit vector pointing on the direction of ${\bf{r}}_{\perp}$.

We notice that the electric field (\ref{eleccline}) is a new effect which appears due to the Lorentz symmetry breaking, if ${\bf{{V}}_{\perp}}=0$ this effect  disappears. The electric field modulus reads
\begin{eqnarray}
\label{emodclll}
\mid{\Delta\bf{E}}\left({\bf{r}}_{\perp}\right)\mid =\frac{\mid I\mid\mid{\bf {V}}_{\perp}\mid}{\pi{\bf{r}}_{\perp}^{2}} \ ,
\end{eqnarray}
which does not depend on the direction of the vector ${\bf{r}}_{\perp}$ with respect to background vector. 

Again, considering an atomic system and $\mid I\mid\sim 10^{5}$A, we can find the following estimates in order of magnitude for the electric field (\ref{emodclll}), $\mid{\Delta\bf{E}}\mid\sim\left(10^{-14}-10^{-20}\right)$N/C (radiative corrections) and $\mid{\Delta\bf{E}}\mid\sim 10^{-24}$N/C (astrophysical birefringence).

The magnetic field generated outside of steady current line can be obtained from Eq. (\ref{AVcurreline}), resulting in
\begin{eqnarray}
\label{bcurrenlinee}
{\bf{B}}\left({\bf{r}}_{\perp}\right)=\frac{I}{2\pi{\bf{r}}_{\perp}^{2}}\Biggl\{\left[1-\frac{8}{{\bf{r}}_{\perp}^{2}}\left(\frac{\left({\bf {V}}_{\perp}\cdot{\bf {r}}_{\perp}\right)^{2}}{{\bf{r}}_{\perp}^{2}}-{\bf {V}}_{\perp}^{2}\right)\right]\left({\bf {r}}_{\perp}\times{\hat{z}}\right)+16\frac{\left({\bf {V}}_{\perp}\cdot{\bf {r}}_{\perp}\right)}{{\bf{r}}_{\perp}^{2}}\left({\bf {V}}_{\perp}\times{\hat{z}}\right)\Biggr\} \ .
\end{eqnarray}

\begin{figure}[!h]
\centering \includegraphics[scale=0.35]{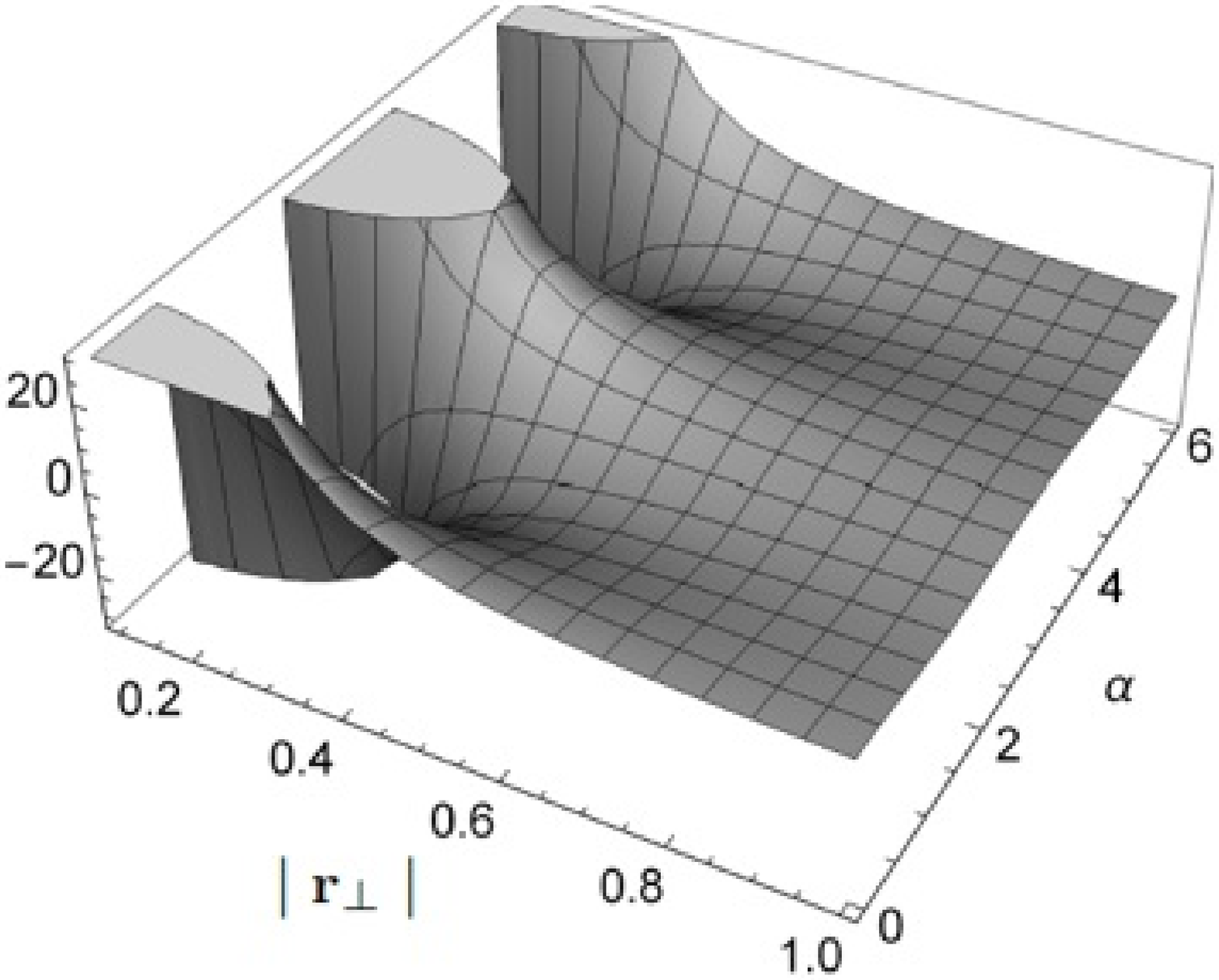} \caption{Function $g\left(\mid{\bf{r}}_{\perp}\mid , \alpha\right)$, appearing in (\ref{gralphacl}).}
\label{mcline}
\end{figure}

The first contribution on the right hand side of Eq. (\ref{bcurrenlinee}) stands for the magnetic field obtained in Maxwell theory, the remaining terms are Lorentz violating corrections of the second order in the background vector.  The modulus of the magnetic is given by
\begin{eqnarray}
\label{bculinemodu1}
\mid{\bf{B}}\left({\bf{r}}_{\perp}\right)\mid=\frac{\mid I\mid}{2\pi\mid{\bf{r}}_{\perp}\mid}-\frac{4\mid I\mid{\bf {V}}_{\perp}^{2}}{\pi}g\left(\mid{\bf{r}}_{\perp}\mid , \alpha\right) \ ,
\end{eqnarray}
where $\alpha\in\left[0,2\pi\right)$ means the angle between ${\bf {V}}_{\perp}$
and ${\bf {r}}_{\perp}$, and we defined the function
\begin{eqnarray}
\label{gralphacl}
g\left(\mid{\bf{r}}_{\perp}\mid , \alpha\right)=\frac{1}{\mid{\bf{r}}_{\perp}\mid^{3}}\left[2\cos^{2}\left(\alpha\right)-1\right] \ ,
\end{eqnarray}
that describes the behaviour of the Lorentz violating correction to the magnetic field founded in Maxwell electrodynamics, as showed in Eq. (\ref{bculinemodu1}). In Fig. \ref{mcline} we have a plot for $g\left(\mid{\bf{r}}_{\perp}\mid , \alpha\right)$.

Taking into account an atomic system e by using $\mid I\mid\sim 10^{5}$A, we can perform an estimate in order of magnitude for the magnetic field produced outside of a steady current line in Maxwell electrodynamics (first contribution on the right hand side of Eq. (\ref{bculinemodu1})), namely $\sim 10^{8}$T.  For the Lorentz violating correction (second term in Eq. (\ref{bculinemodu1})), we obtain the estimates: $\mid{\Delta\bf{B}}\mid\sim\left(10^{-53}-10^{-65}\right)$T (radiative corrections) and $\mid{\Delta\bf{B}}\mid\sim 10^{-73}$T (astrophysical birefringence).

As final comment we point out that by following the same steps as in Subsect. \ref{V}, we can show that the presence of Dirac strings do not produce any obvious effect up to second order in $V^{\mu}$ for the nonminimal Lorentz violating model (\ref{lagEmNM}).

\section{Conclusions and final remarks}

\label{conclusoes} 

In this paper the interactions between stationary electromagnetic
sources for the minimal and nonminimal CPT-odd photon sector of the
SME were investigated. We have obtained perturbative results up to
second order in the Lorentz breaking parameters and focused mainly
on physical phenomena that have no counterpart in Maxwell theory.

First, we considered the Maxwell electrodynamics modified by the CFJ
term, which belongs to the minimal SME. We have shown the emergence
of a spontaneous torque on a classical dipole and the occurrence of
an electromagnetic filed produced by a static point-like 
charge. We showed that the Dirac string interacts with a point-like
charge and with another string. We also calculated
the electromagnetic field produced outside a Dirac string and investigated
the Lorentz violation modifications on the Aharonov-Bohm bound states.

After, we considered a model which is the higher-derivative version
of the CFJ one, where the Lorentz violating term has dimension five
and belongs to the CPT-odd sector of the nonminimal SME. We showed
that the presence of point-like charges and a steady current line
induce nontrivial effects in low energy physics as spontaneous torques and electromagnetic fields. 

We also performed
some numerical estimates in order to investigate the relevance of
some of the obtained physical phenomena in atomic systems. We also have made an overestimate for the background vectors using experimental data from the atomic electric field.

\begin{acknowledgments}
This study was financed in part by the Coordena\c c\~ao de Aperfei\c coamento de Pessoal de N\'\i vel
Superior -- Brasil (CAPES) -- Finance Code 001 (LHC). AFF acknowledges the support by Conselho Nacional de
Desenvolvimento Cient\'\i fico e Tecnol\'ogico (CNPq) via the grant 305967/2020-7.
\end{acknowledgments}


\end{document}